\documentclass[10pt,a4paper]{article}
\usepackage{physmtemplate-article}

\begin{document}


\title{Active remodeling rescues a microswimmer from geometric trapping}
\author[1]{Ayushi Bhatt}
\author[2]{Nico Schramma}
\author[1]{Jasper van der Gucht}
\author[1]{Uddalok Sen\footnote{uddalok.sen@wur.nl, ORCID: \href{https://orcid.org/0000-0001-6355-7605}{0000-0001-6355-7605}}}
\affil[1]{Physical Chemistry and Soft Matter Group, Wageningen University and Research, 6708 WE Wageningen, The Netherlands}
\affil[2]{Department of Medical Biochemistry, Amsterdam UMC, University of Amsterdam, Meibergdreef 9, 1105 AZ Amsterdam, The Netherlands}
\date{}
\maketitle


\begin{abstract}
    Natural habitats confront microorganisms with heterogeneous, mechanically complex surroundings that shape how they move and explore. Most studies of motility in complex media treat the environment as either a homogeneous fluid or a rigid obstacle field. Here we instead study a medium that is simultaneously soft and reconfigurable --- deformable by the organism moving through it. We track the unicellular biflagellate microalga \emph{Chlamydomonas reinhardtii} swimming in a soft, disordered suspension of microgel particles comparable in size to the cell itself, across a range of gel packing fractions. With increasing packing fraction, trajectories change from long straight runs to short, frequently reorienting ones, and the effective translational diffusivity decreases monotonically. The bimodal run/tumble speed distribution, characteristic of the free-swimming gait, collapses toward a single mode as the packing fraction increases, by near-proportional slowing of both gait states rather than a redistribution of cells between them. Recasting trajectories as alternating hops and traps reveals the signature of a deformable medium: hop lengths halve as the gel densifies, approaching the microgel aggregate scale, while trapping times do not increase with increasing packing fraction. Independent force estimates place the cell's propulsive force close to an order of magnitude above that needed to deform a single grain, so remodeling is expected rather than merely observed. A minimal \emph{active remodeler} model --- an active Brownian particle among mobile, rigid obstacles --- shows that obstacle mobility alone is sufficient to relocate the onset of permanent trapping. Thus, our results establish obstacle mobility as a control parameter for motility in soft, disordered matter. 
\end{abstract}





\section{Introduction}

Flagellar propulsion is among the earliest and most widespread forms of microbial motility \cite{cosson-1996-cellbiolint, liu-2007-pnas}. It underlies processes ranging from infection and biofilm formation \cite{costerton-1999-science, josenhans-2002-intjmedmicrobiol} to plankton transport in lakes and oceans \cite{park-2001-phycologia,durham-2013-natcommun}, and the biodegradation of pollutants \cite{valentine-2010-science}. Unicellular swimmers --- bacteria such as \emph{Escherichia coli}, sperm cells, and algae such as \emph{Chlamydomonas reinhardtii} --- operate at low Reynolds numbers, where inertia is irrelevant and macroscopic swimming strategies fail \cite{purcell-1977-amjphys, lauga-2009-rpp, guasto-2012-arfm, elgeti-2015-rpp}, and whose behavior in complex, structured fluids is the subject of a growing literature \cite{araujo-2023-sm, spagnolie-2023-arcmp, gompper-2025-jphys}. \emph{C. reinhardtii} --- a unicellular, biflagellate green alga that swims flagella-first with a breaststroke gait --- has become a canonical model for eukaryotic microswimmer physics \cite{drescher-2010-prl, guasto-2010-prl, goldstein-2015-arfm, jeanneret-2016-epjspectop}. In the dark, it executes a eukaryotic analog of run-and-tumble motility, stochastically switching between synchronous beating that produces near-straight runs and asynchronous beating that produces abrupt reorientations \cite{polin-2009-science, wan-2019-ptrsb, wei-2024-elife} --- a gait structure whose distinct behavioral states can be resolved by long-term single-cell phenotyping in confinement \cite{bentley-2022-elife}. \\

In their natural settings, however, cells rarely swim through simple fluids. They migrate through the dense extracellular matrix of tissues \cite{pathak-2012-pnas}, the labyrinthine pore space of soil \cite{ranjard-2001-resmicrobiol}, and viscoelastic mucus \cite{ottemann-1997-molmicrobiol}; swimming algae in particular disperse through structured and porous habitats \cite{croze-2013-jrsi, thery-2021-sm, souzy-2022-prr, lagoin-2025-pnas}. Environmental complexity can be rheological, geometric, or mechanical, and each has been studied largely in isolation. Rheologically, polymer additives can \emph{enhance} bacterial motility by suppressing tumbling and reducing body wobble \cite{martinez-2014-pnas, patteson-2015-scirep, zottl-2019-natphys, kamdar-2022-nature}, yet the same class of additives \emph{hinders} the swimming of \emph{C. reinhardtii}, whose breaststroke motion couples strongly to the fluid elasticity \cite{qin-2015-scirep}. Geometrically, \emph{C. reinhardtii} scatters from flat and curved boundaries primarily through ciliary contact, reorienting on contact and leaving walls at shallow angles \cite{kantsler-2013-pnas, contino-2015-prl, thery-2021-sm, bruncosmebruny-2026-arxiv}, and curved confinement restricts its exploration to the boundary geometry \cite{ostapenko-2018-prl, cammann-2021-pnas, till-2022-prr, lagoin-2025-pnas, laroussi-2026-small}. Long-range hydrodynamic interactions with surfaces likewise produce hopping trajectories --- repeated excursions away from and returns to the surface --- that govern where cells accumulate \cite{buchner-2021-pnas}. Mechanically disordered media add a further ingredient: in rigid, disordered pore spaces, bacterial run-and-tumble is replaced by intermittent \emph{hopping-and-trapping}, with broadly distributed hop lengths and trapping times that nonetheless yield long-time diffusion \cite{bhattacharjee-2019-natcommun, bhattacharjee-2019-sm}. In periodic pillar arrays, increasing obstacle density reduces both the directional correlation time and the effective diffusivity, suppressing long-range exploration \cite{bruncosmebruny-2019-jchemphys}. \\

Two features of real biological matrices are not yet captured by these simplified model systems. First, many are \emph{structured at the scale of the swimmer itself}: rather than a homogeneous continuum, the cell meets discrete, deformable elements --- polymer grains, aggregates, mesh strands --- comparable in size to its own body. Whether such a medium acts as an effective continuum or as a discrete, heterogeneous landscape is set by the ratio, $\alpha$, of the microstructural grain size to the swimmer size (a scale separation familiar from the microrheology of soft, structured materials \cite{squires-2010-arfm}). Recent work on bacteria in microgel suspensions shows that microscale grain disorder reshapes exploration, producing a ``medium-assisted tumbling'' regime governed by \emph{local}  mechanical resistance rather than intrinsic navigation \cite{urra-2025-sm}; we adopt the same class of medium in this work. Second, and most importantly, real matrices are \emph{reconfigurable}: the swimmer can deform, displace, and locally restructure them. The paradigmatic example is \emph{Helicobacter pylori}, which chemically liquefies gastric mucin to swim through it \cite{celli-2009-pnas}; more generally, active particles are known to distort and even destroy the disordered landscapes they traverse \cite{morin-2017-natphys}. A medium that is simultaneously soft, disordered at the swimmer scale ($\alpha \approx 1$), \emph{and} remodelable therefore sits outside both the ``homogeneous complex fluid'' and the ``rigid obstacle field'' configurations that dominate the literature. \\

Concretely, this size ratio $\alpha$ sorts microswimmers into three regimes. A microswimmer much larger than the grains ($\alpha \ll 1$) averages over many of them and feels an effective bulk yield stress, so locomotion reduces to the classical problem of swimming in a viscoplastic fluid \cite{hewitt-2017-jfm, supekar-2020-jfm, nazari-2023-prl}. Bacteria such as \emph{E. coli} (body size $\sim$ 1 \SI{}{\text{\micro\meter}}) are far smaller than the grains they encounter ($\alpha \gg 1$), and thread the pore space; their intermittent, stop-and-go motion set by the local resistance to yielding at constrictions \cite{bhattacharjee-2019-natcommun, urra-2025-sm}. In our experiments, \emph{C. reinhardtii} sits in neither limit: its body size ($\sim$ 10 \SI{}{\text{\micro\meter}}) is comparable to individual microgel grains and smaller than the aggregates it encounters; so, it neither averages over a continuum nor threads a static pore network. Instead, it potentially deforms and displaces individual grains and clumps locally in a dense, disordered packing \cite{seth-2011-natmater, nicolas-2018-rmp}. This intermediate regime ($\alpha \approx$ 1), however, remains largely unexplored.  \\

Here we bridge this gap. We study \emph{C. reinhardtii} swimming in Carbopol microgel suspensions across effective volumetric packing fractions from mild to dense crowding. This medium is soft, disordered at the scale of the cell, and --- unlike rigid pillar arrays or fixed-bead packings --- the grains comprising the medium can be deformed and displaced by the swimmer (as we show below). Combining single-cell tracking, direct confocal visualization of the microgel suspension, and a minimal mobile-obstacle model, we show that the effective confinement a cell experiences is dynamically coupled to its own activity, and that the \emph{mobility} of the surroundings is the parameter that separates sustained transport from permanent geometric trapping.  

\section{Results}

\subsection{Soft confinement progressively suppresses spatial exploration}

\begin{figure}
    \centering
    \includegraphics[width=\textwidth]{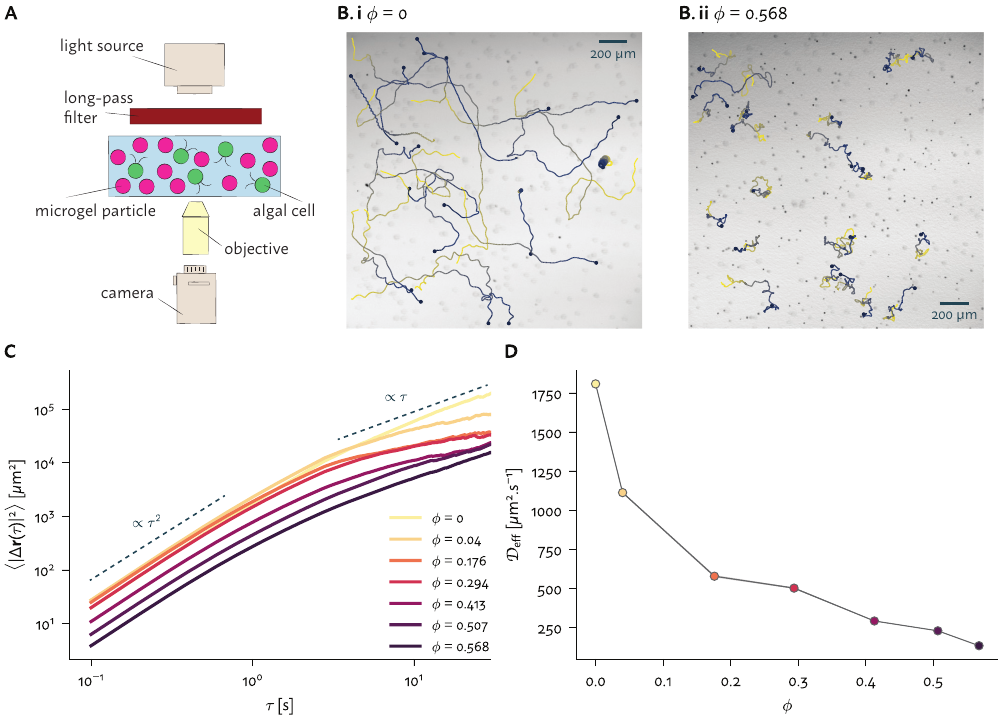}
    \caption{\textbf{Soft, disordered confinement progressively suppresses spatial exploration.} \textbf{A.} Schematic of the experimental setup: motile \emph{C. reinhardtii} algal cells in a microgel suspension, imaged in a quasi-two-dimensional chamber under long-pass-filtered illumination to avoid phototactic bias. Representative trajectories of motile cells \textbf{B.i.} at $\phi$ = 0, showing extended, weakly curved paths, and \textbf{B.ii.} at $\phi$ = 0.568, where trajectories become markedly more tortuous and spatially confined. Trajectory colors encode elapsed time, increasing monotonically along each path. See movie SM1 for the corresponding experimental movies. \textbf{C.} Ensemble mean-squared displacement (MSD) $\langle \vert \Delta \mathbf{r} (\tau) \vert^{2} \rangle$ versus lag time $\tau$ for each volume fraction $\phi$, showing the ballistic-to-diffusive crossover, and its systematic shift to shorter lag times with increasing $\phi$. \textbf{D.} Effective translational diffusivity $\mathcal{D}_{\mathrm{eff}}$ decreases monotonically with $\phi$, indicating a reduced spatial exploration of the suspension at higher microgel densities. }
    \label{fig:fig-1}
\end{figure}

\emph{C. reinhardtii} is a unicellular, biflagellate green alga whose free-swimming motility follows a eukaryotic run-and-tumble gait: cells alternate between directed runs, during which the two flagella beat in a synchronized breaststroke, and brief tumbles, during which flagellar coordination is transiently lost and the cell reorients \cite{polin-2009-science}. To quantify how soft, heterogeneous confinement induced by a microgel dispersion modifies this motility, we optically tracked individual cells in a quasi-two-dimensional observation chamber across a range of Carbopol effective volume fractions, $\phi$ (figure \ref{fig:fig-1}A; see Methods for experimental details and figure \ref{fig:fig-s1}B for volume fraction assignment). All recordings were acquired in a single focal plane; the analysis is based on the resulting planar trajectories (see figure \ref{fig:fig-s2} for a discussion on the validity of the planar analysis). \\

In Tris-acetate-phosphate (TAP; see Methods for experimental details) growth medium ($\phi = 0$), the cells display their characteristic gait, with extended, weakly curved runs punctuated by reorientation events (figure \ref{fig:fig-1}B.i and movie SM1). At the highest packing fraction studied ($\phi$ = 0.568), trajectories become markedly more tortuous and spatially confined (figure \ref{fig:fig-1}B.ii and movie SM1). From these trajectories, we computed the ensemble mean-squared displacement (MSD), $\langle \vert \Delta \mathbf{r} (\tau) \vert^{2} \rangle$, as a function of lag time $\tau$. In TAP, the MSD crosses over from ballistic scaling ($\langle \vert \Delta \mathbf{r} (\tau) \vert^{2} \rangle \propto \tau^{2}$) at short times to diffusive scaling ($\langle \vert \Delta \mathbf{r} (\tau) \vert^{2} \rangle \propto \tau$) at long times --- the hallmark of run-and-tumble dynamics \cite{howse-2007-prl, tenhagen-2011-jphys} (figure \ref{fig:fig-1}C). With increasing $\phi$, this crossover occurs at systematically smaller displacements. \\

The consequence for exploration is a monotonic loss of effective diffusivity $\mathcal{D}_{\mathrm{eff}}$ --- extracted from the long-time (diffusive) slope of the MSD --- across $\phi$ (figure \ref{fig:fig-1}D). A monotonic decline of the effective diffusivity is the generic response of a self-propelled particle to a crowded environment \cite{bechinger-2016-rmp}, reported both for synthetic active colloids and for cells moving through fixed obstacle arrays \cite{morin-2017-natphys, zeitz-2017-epje, bruncosmebruny-2019-jchemphys}. Notably, a falling effective diffusivity with increasing obstacle density has been reported for the same organism swimming through \emph{rigid}, fixed pillar lattices \cite{bruncosmebruny-2019-jchemphys}. This bulk transport observable therefore does not by itself distinguish a medium that yields to the swimmer from one that does not --- a distinction that requires the trajectory-resolved analysis we turn to next.  

\subsection{The run-tumble gait collapses and directional persistence decays}

\begin{figure}
    \centering
    \includegraphics[width=\textwidth]{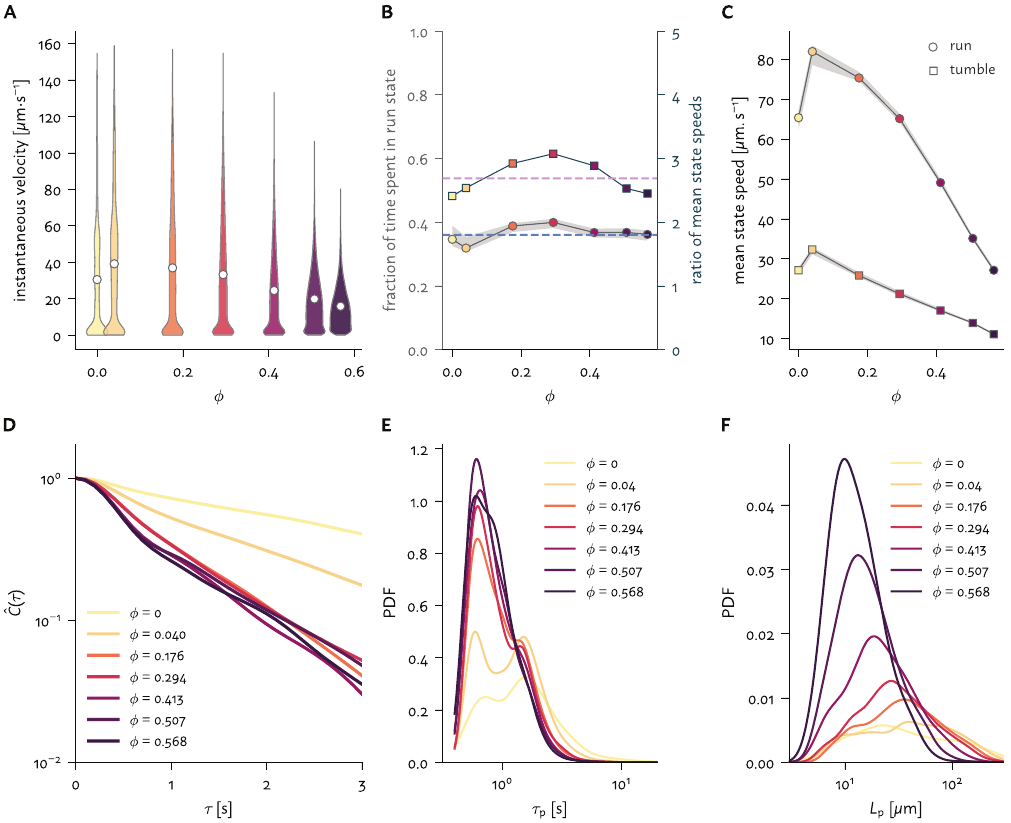}
    \caption{\textbf{The gait bimodality collapses by proportional slowing of both run and tumble states, while directional persistence decays faster and becomes more uniform.} \textbf{A.} Distributions of instantaneous swimming speed at each volume fraction $\phi$; the distribution narrows and its bimodal structure fades with increasing $\phi$. Overlaid circular markers denote the mean of each distribution. \textbf{B.} Fraction of time spent in the run state (circular markers; left axis) remains unchanged against $\phi$ (dashed line: mean of 0.36), as well as the ratio of mean run and tumble state speeds (square markers; right axis; dashed line: mean of 2.69), showing that the collapse of bimodality is not a redistribution of cells between the gait states; shaded region denotes a 95\% confidence interval. \textbf{C.} Mean speeds of the run (circular markers) and tumble (square markers) states versus $\phi$; shaded region denotes a 95\% confidence interval. \textbf{D.} Normalized velocity autocorrelation $\hat{C} (\tau)$ versus lag time $\tau$; the decay accelerates with increasing $\phi$. Distributions of the per-trajectory \textbf{E.} persistence time $\tau_{\mathrm{p}}$ --- which transitions from bimodal to unimodal with increasing $\phi$ --- and \textbf{F.} persistence length $L_{\mathrm{p}}$, with the latter markedly shifting toward smaller values with increasing $\phi$, reflecting a progressive loss of directional memory at higher microgel densities. }
    \label{fig:fig-2}
\end{figure}

The instantaneous speed distribution in TAP medium ($\phi$ = 0) is distinctly multimodal (figure \ref{fig:fig-2}A): a low-speed population corresponding to tumbles and pauses coexists with a high-speed run population \cite{polin-2009-science}. Multimodality is highly significant in TAP ($\phi$ = 0) and in the sparsest gel ($\phi$ = 0.04), but is no longer statistically detectable at intermediate and high $\phi$ (figures \ref{fig:fig-2}A and \ref{fig:fig-s3}A). Run and tumble states are isolated from the pooled speed distribution at each $\phi$ using a two-component Gaussian mixture model (see Methods); state mean speeds and the fraction of time spent running are then computed from this classification (figures \ref{fig:fig-2}B and \ref{fig:fig-2}C). The fraction of time spent in the run state remains virtually unchanged across the entire $\phi$ range (circular markers and left axis in figure \ref{fig:fig-2}B; dashed line, left axis: mean fraction of 0.36), so the collapse is not a redistribution of cells between running and tumbling states. Instead, both state speeds decrease with increasing $\phi$ (figure \ref{fig:fig-2}C), and their ratio stays close to 2.69 (dashed line, right axis) across the entire range studied (square markers and right axis in figure \ref{fig:fig-2}B) --- essentially preserved rather than collapsing to unity. Hence, the gel does not erode the relative contrast between the gait states; it slows both proportionally, so that the absolute speed gap shrinks below the width of the distribution and the two modes merge. This is consistent with a medium that increases resistance in both the synchronous and asynchronous beating flagellar modes rather than one selectively suppressing power strokes. The strength of this coupling depends on the details of the flagellar stroke: mutant strains with altered flagellar coordination show systematically different suppression (figures \ref{fig:fig-s7} and \ref{fig:fig-s8}), confirming that the swimmer-medium coupling is set by the gait and not by cell size alone.  \\

Notably, this suppression of speed is not monotonic in $\phi$. The mean instantaneous speed peaks in the sparsest gel ($\phi$ = 0.04, not $\phi$ = 0), and only at higher $\phi$ values does it fall below its free-swimming value (figures \ref{fig:fig-2}C and \ref{fig:fig-s3}B). To ask whether this enhancement originates in the discrete microgel packing or in the continuous phase between the grains, we centrifuged the microgel suspension and tracked cells in the resulting grain-free supernatant (figure \ref{fig:fig-s5}). The supernatant viscosity rises with $\phi$ (figure \ref{fig:fig-s5}A), indicating a population of dissolved, uncrosslinked Carbopol polymer in the continuous phase, and cells swimming in it reproduce the same speed increase between $\phi$ = 0 and $\phi$ = 0.04 seen in the complete gel (figures \ref{fig:fig-2}B, \ref{fig:fig-s3}B, and \ref{fig:fig-s5}B). The enhancement is therefore carried by the dissolved-polymer phase rather than by the grain packing itself. Whether fluid elasticity (due to the presence of dissolved polymers) raises or lowers swimming speed is known to depend sensitively on the swimmer's gait, stroke amplitude, and body flexibility \cite{teran-2010-prl, thomases-2014-prl}. A modest speed enhancement on adding polymers is reminiscent of that reported for flagellated bacteria in polymer solutions, where suppressed wobbling and reduced body drag can outweigh the added viscosity \cite{martinez-2014-pnas, patteson-2015-scirep, zottl-2019-natphys, kamdar-2022-nature}. However, the flow fields of bacterial pushers and algal pullers differ enough that the mechanism need not transfer directly \cite{book-lauga}. For \emph{C. reinhardtii} specifically, dilute elastic polymer solutions have been found to \emph{hinder} net swimming despite raising the flagellar beat frequency \cite{qin-2015-scirep}. Our supernatant measurements (figure \ref{fig:fig-s5}A) resolve a rising viscosity but the associated elastic moduli were too small to quantify; so we identify the dissolved, uncrosslinked Carbopol polymer as the agent of the enhancement without assigning the mechanism to a specific viscous or elastic contribution. \\ 

We now turn to why $\mathcal{D}_{\mathrm{eff}}$ nevertheless falls monotonically from the first gel concentration onward (figure \ref{fig:fig-1}D), even though speed itself rises at $\phi$ = 0.04. $\mathcal{D}_{\mathrm{eff}}$ is set jointly by speed and directional persistence (orientation correlation time), which are independent of each other. To ask whether that is the case here, we characterize this directional persistence via the velocity autocorrelation function (VACF) \cite{book-furst},
\begin{align}
    C (\tau) = \langle \mathbf{v} (t) \cdot \mathbf{v} (t + \tau) \rangle ,
\end{align}
and its normalized form $\hat{C} (\tau) = C (\tau) / C(0)$. Fitting \cite{book-furst} $\hat{C} (\tau) = \exp (- \tau / \tau_{\mathrm{p}})$ yields the persistence time $\tau_{\mathrm{p}}$, and with the mean track speed $v_{\mathrm{m}}$, the persistence length $L_{\mathrm{p}} = \tau_{\mathrm{p}} \, v_{\mathrm{m}}$. With increasing $\phi$, the normalized VACF decays more rapidly (figure \ref{fig:fig-2}D), and both $\tau_{\mathrm{p}}$ and $L_{\mathrm{p}}$ distributions shift to smaller values (figures \ref{fig:fig-2}E and \ref{fig:fig-2}F). As with the MSD's ballistic-to-diffusive crossover, the largest change in VACF (figure \ref{fig:fig-2}D) occurs around the sparsest gel concentration ($\phi$ = 0.04). Hence, the initial loss of exploration is attributable to a loss of directional persistence, not of speed. \\

Crucially, the distributions do not merely shift. The $\tau_{\mathrm{p}}$ distribution also changes shape --- it is bimodal at low $\phi$ and becomes unimodal at intermediate $\phi$ (figure \ref{fig:fig-2}E). The $L_{\mathrm{p}}$ distribution, by contrast, narrows prominently with $\phi$ (figure \ref{fig:fig-2}F). This narrowing is not a trivial consequence of the overall slowing: the coefficient of variation of $\tau_{\mathrm{p}}$ declines monotonically with $\phi$ (figure \ref{fig:fig-s3}C); so the distribution narrows in relative as well as in absolute terms. Cells in a dense gel are therefore not only less persistent on average, but also \emph{more alike} each other in how persistent they are --- the broad tail of highly persistent swimmers in free suspension is progressively removed with increasing $\phi$ (figure \ref{fig:fig-2}F). This runs counter to the intuition --- borne out in rigid porous media \cite{chepizhko-2013-prl, bhattacharjee-2019-natcommun} --- that quenched disorder \emph{broadens} the distribution of single-cell transport parameters by trapping some cells while sparing others. That the opposite occurs here is a first hint that the medium in the present experiments is not a fixed landscape: rather than sampling a frozen distribution of local environments, each cell reshapes its own surroundings, so that the escape dynamics converge on a common, swimmer-set scale. \\

Superimposed on this narrowing there is, nonetheless, evidence for dynamically distinct subpopulations at intermediate crowding (figure \ref{fig:fig-2}E). This is consistent with cells sampling distinct local environments --- some comparatively open, others more obstructed. The overall picture is one of a medium that progressively imposes its own length and time scales on the swimmer, overriding the intrinsic variability of the flagellar gait. The transient subpopulations at intermediate crowding likely reflects real heterogeneity in local grain density that narrows away as $\phi$ increases further. Whether this narrowing can be attributed to the medium's intrinsic microstructure, or whether the swimmer actively deforms that microstructure as it moves, is the question we turn to next. 

\subsection{Cells actively deform the microstructure}

\begin{figure}
    \centering
    \includegraphics[width=\textwidth]{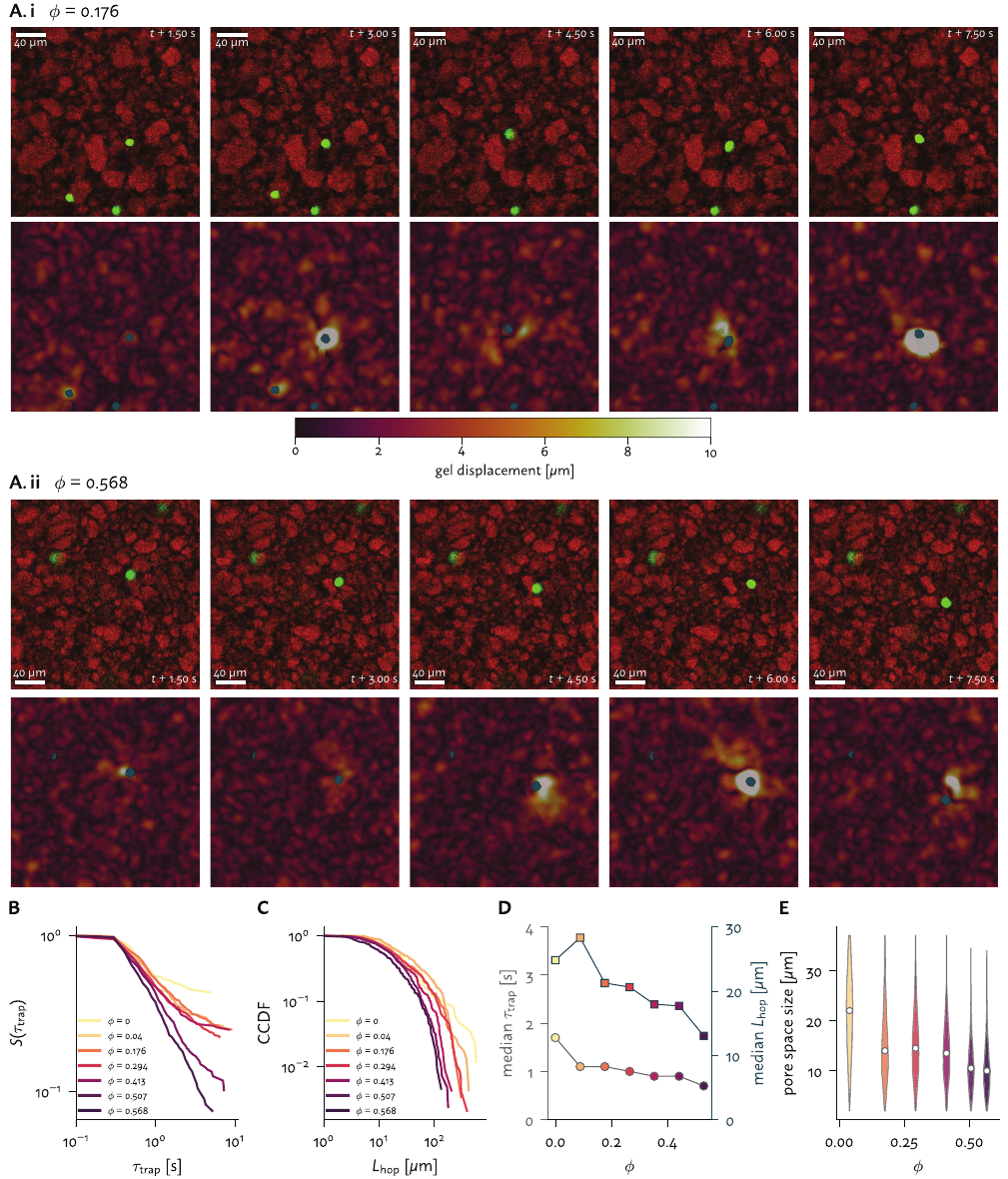}
    \caption{\textbf{Cells actively remodel the microgel and, as a result, take shorter steps without longer waits.} Time-lapsed confocal micrographs of motile cells (green) exploring the microgel network (red) at \textbf{A.i.} $\phi$ = 0.176 and \textbf{A.ii.} $\phi$ = 0.568. At both volume fractions the local microgel structure is visibly displaced and reorganized around the advancing swimmer and relaxes back once the cell has passed, indicating that the gel remains locally compliant even at the highest packing fractions studied (upper row of each panel). See movie SM2 for the corresponding experimental movies. Below each panel, the corresponding displacement colormap (frame-to-frame, i.e.\ instantaneous; see Methods) renders the same field of view as a map of local gel displacement magnitude, with the cell itself shown as a fixed-color mask. \textbf{B.} Kaplan-Meier survival function $S (\tau_{\mathrm{trap}})$ of trapping times $\tau_{\mathrm{trap}}$ and \textbf{C.} Complementary cumulative distribution (CCDF) of hop lengths $L_{\mathrm{hop}}$ for different $\phi$. \textbf{D.} Median trapping time (circular markers; left axis) and median hop length (square markers; right axis) versus $\phi$. The hop length halves while trapping times do not increase. \textbf{E.} Pore space-size distribution of the labeled microgel network versus $\phi$, obtained by granulometry (morphological opening of the void space with circular disks of increasing radius; see Methods); overlaid circular markers denote the median of each distribution. }
    \label{fig:fig-3}
\end{figure}

The analysis above suggests that microstructural rearrangements induced by the microswimmer might be responsible for the observed dynamics. We investigate this using time-lapsed dual-channel fluorescence microscopy of the autofluorescent motile cell within the fluorescently-labeled microgel dispersion (see Methods for experimental details). In both sparse ($\phi$ = 0.176; upper row of figure \ref{fig:fig-3}A.i and movie SM2) and dense ($\phi$ = 0.568; upper row of figure \ref{fig:fig-3}A.ii and movie SM2) gels, the local microgel structure visibly deforms and is displaced around the advancing cell, before (partially) returning toward its original configuration once the cell has passed. The displacement colormaps computed from the same recordings (lower rows of figures \ref{fig:fig-3}A.i and \ref{fig:fig-3}A.ii; see Methods) makes this pattern spatially explicit: a localized region of elevated displacement tracks the cell's advancing front and fades again once the cell has passed, at both $\phi$ = 0.176 and $\phi$ = 0.568 --- a direct, quantitative rendering of the deform-and-recover cycle the raw fluorescence channel only shows qualitatively. The medium is thus locally compliant enough to be pushed aside by flagellar forces at every concentration studied, and the post-contact recovery is consistent with the soft, elastic character of individual microgel grains (the cell remains in contact with the microgel grain for $\approx$ 1.5 \SI{}{\text{\second}} while post-contact recovery of the grain spans $\approx$ 4.5 \SI{}{\text{\second}}). Active reshaping of a disordered environment is not unique to this system --- \emph{C. reinhardtii} itself measurably stiffens a jammed emulsion of oil droplets it swims through via confinement-amplified propulsive forces at the droplet contacts \cite{corpart-2026-arxiv}. Here, remodeling of the medium is achieved by a swimmer displacing individual grains that then recover. A key observation is that remodeling persists into the densest gel. It shows that the swimmer's mechanical influence on its surroundings does not switch off as the packing fraction reaches the highest values we studied --- precisely the regime in which a rigid medium would be expected to arrest the cell permanently. This active remodeling is conceptually distinct from confinement by rigid, immovable obstacles \cite{bhattacharjee-2019-natcommun, bruncosmebruny-2019-jchemphys, sinaasappel-2025-prl}, and it makes a testable kinematic prediction, which we now examine. \\

The confocal micrographs show that cells displace the surrounding medium; they do not show whether this has any influence on how they explore the medium. For bacterial motion in rigid porous media, there is no run-and-tumble but intermittent \emph{hopping and trapping}, in which cells are transiently arrested in pores and released by rare fluctuations, producing hop lengths set by the pore geometry and broadly, often heavy-tailed, distributed trapping times \cite{bhattacharjee-2019-natcommun, bhattacharjee-2019-sm, sinaasappel-2025-prl}. Whether a microswimmer penetrates such a medium at all is governed by geometric criteria relating its size and persistence to the pore-space statistics \cite{kurzthaler-2021-natcommun}. So, a natural expectation for a denser medium is longer and more broadly distributed traps. To ask whether \emph{C. reinhardtii} in a deformable gel behaves the same way, we segmented every trajectory between alternating hops (instantaneous speed above a threshold) and traps (below it), using a threshold relative to the median speed of each condition so that the segmentation is not biased by the shifting speed distribution (see Methods, and figure \ref{fig:fig-s4} for sensitivity on threshold choice). Trapping-time distributions were estimated by the Kaplan-Meier method \cite{kaplan-1958-jasa}, which accounts for traps truncated by the end of the track (figures \ref{fig:fig-3}B and \ref{fig:fig-3}C). \\

Two results emerge. First, the hop length halves as the gel packing fraction increases (figure \ref{fig:fig-3}C, and square markers and right axis in figure \ref{fig:fig-3}D) --- approaching the scale of the cell body and of the microgel aggregates it encounters (figure \ref{fig:fig-3}E). The medium therefore geometrically limits how far a cell travels before its next encounter --- the deformable-medium counterpart of the pore-set hop length in rigid porous media \cite{bhattacharjee-2019-sm}. Second, and in sharp contrast to the rigid-porous-medium picture, trapping times do not increase (figure \ref{fig:fig-3}B, circular markers and left axis in figure \ref{fig:fig-3}D, and figure \ref{fig:fig-s4}A). Since this is a claim about what fails to happen, it needs a quantitative benchmark. To that end, we construct a null test corresponding to a non-deformable medium: each $\phi$ = 0 cell's own measured path, replayed at the slower speed measured for each non-zero $\phi$ (figure \ref{fig:fig-s8}A). This null predicts trapping times that increase with crowding, since a medium that merely impedes the cell's motion without deforming should trap a slower cell for longer. The measured trend runs in the opposite direction to the null test at every packing fraction (figure \ref{fig:fig-s8}A). This is the opposite of the rigid-porous-medium expectation. \\

This combination --- shorter steps but not longer waits --- is the kinematic signature of a medium that is deformed by the swimmer. In a rigid porous medium, a cell that wanders into a trapping cage must wait for a fluctuation to release it; a slower cell in a tighter, fixed cage should therefore show pronounced lengthening of trapping times, especially since our cells also slow roughly threefold across the concentration series (figure \ref{fig:fig-2}C). We, however, observe no such lengthening in our experiments. The confinement a cell experiences must therefore be dynamically coupled to its own activity: the cell does not wait to be released, because it displaces the obstacles itself (exactly as figures \ref{fig:fig-3}A.i and \ref{fig:fig-3}A.ii, and movie SM2 show). That is also why the persistence-time distribution narrows and loses its bimodal shape with increasing $\phi$ (figure \ref{fig:fig-2}E): escape is governed by the swimmer's own stroke, which is far more reproducible from cell to cell than the disordered pore geometry would be. \\

Finally, the relevant length scales are comparable. The characteristic pore-space size-distribution of the labeled microgel dispersion was quantified by granulometry \cite{book-matheron, book-serra, book-soille} --- the diameter of the largest disk that fits within the void space, which sets whether a cell of a given size can occupy a region at all (figure \ref{fig:fig-3}E; see Methods for experimental details). The median of this distribution falls with $\phi$ and approaches 10 \SI{}{\text{\micro\meter}} at the highest concentration, comparable to the cell size ($\approx$ 10 \SI{}{\text{\micro\meter}}, $\alpha \approx$ 1) and of the same order as the measured median hop length of 13 \SI{}{\text{}{\micro\meter}} (figure \ref{fig:fig-3}D, right axis). The cell is therefore neither much larger than the microstructure (as assumed by a continuum yield-stress description; the $\alpha \ll$ 1 limit) nor much smaller (as for bacteria threading a pore space; the $\alpha \gg$ 1 limit), but of the same order --- confirming that a hop is set by the spacing between the obstacles the cell must displace. 

\subsection{Obstacle mobility preserves exploration that a fixed medium would suppress}

\begin{figure}
    \centering
    \includegraphics[width=\textwidth]{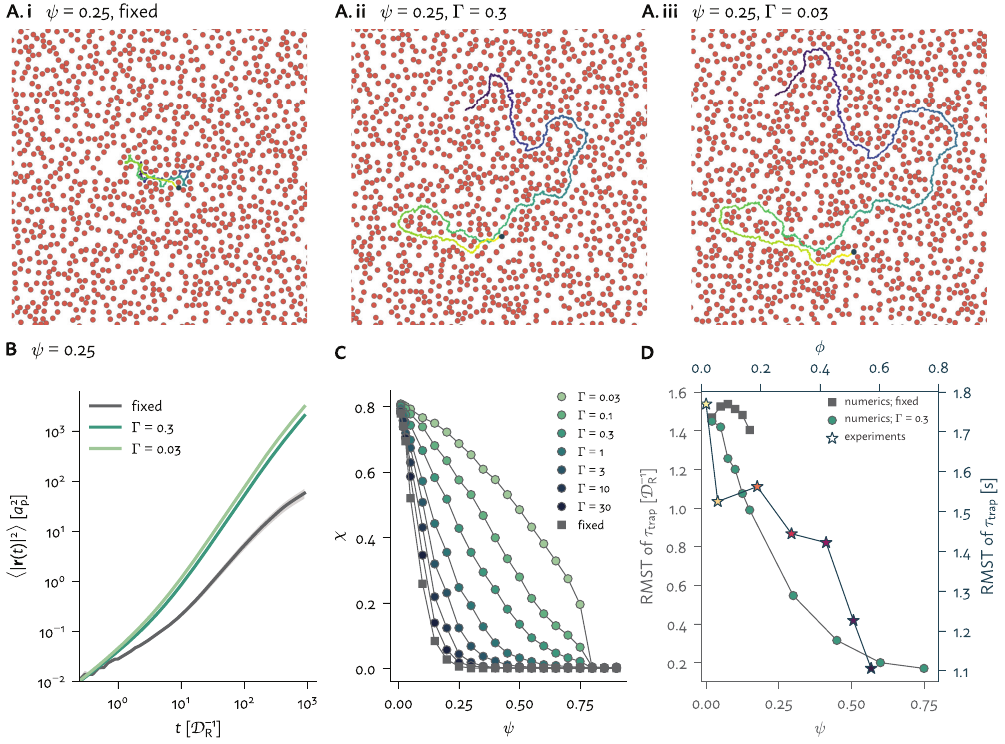}
    \caption{\textbf{Obstacle mobility preserves exploration and rescues the swimmer from the confinement-driven arrest a fixed medium would otherwise impose.} Representative trajectories, from numerical simulations, at obstacle area fraction $\psi$ = 0.25, with the same disordered obstacle field (red disks), for \textbf{A.i.} fixed obstacles, and mobile obstacles with \textbf{A.ii.} $\Gamma$ = 0.3 and \textbf{A.iii.} $\Gamma$ = 0.03. Trajectory color encodes elapsed time, increasing monotonically along each path. See movie SM3 for the corresponding simulation movies. \textbf{B.} Ensemble mean-squared displacement for the conditions in panels A.i--A.iii at $\psi$ = 0.25; shaded bands are the ensemble standard error. \textbf{C.} Normalized exploration $\chi = \mathcal{D}_{\mathrm{eff}} / \mathcal{D}_{\mathrm{free}}$ versus obstacle area fraction $\psi$, shown for a representative set of mobility ratios $\Gamma$ spanning the fixed limit to the most mobile case simulated. The fixed-obstacle curve (square markers) collapses to complete arrest by $\psi \approx$ 0.25 and is the lower envelope of the entire family; mobile media (circular markers) hold $\chi$ high to far higher densities, until all curves converge near $\psi \approx$ 0.78 where the obstacle network itself jams. \textbf{D.} Trap-time ($\tau_{\mathrm{trap}}$) restricted mean survival time (RMST) versus $\psi$ for representative mobile ($\Gamma$ = 0.3; circular markers) and fixed (square markers) cases (left and bottom axes) from the numerical simulations, and versus $\phi$ for the experimental measurements (pentagram markers; right and top axes). The fixed data series is shown only where defined --- every realization at denser $\psi$ is permanently arrested by the end of the observation window (see Methods), leaving no motile subset from which to compute RMST. $\psi$ and $\phi$ are not the same physical quantity; the two axes are aligned for comparison of trend only (see Methods). }
    \label{fig:fig-4}
\end{figure}

The confocal micrographs show the microgel dispersion being both deformed and pushed aside as cells pass (figures \ref{fig:fig-3}A.i and \ref{fig:fig-3}A.ii), and the two responses are difficult to separate experimentally. We therefore ask, in a minimal model, whether the simpler of the two --- bodily displacement of grains, with no change in grain shape --- is on its own sufficient to relieve the suppression that a fixed porous medium imposes. The model deliberately omits grain compliance (grain deformability): obstacles are rigid disks that can only translate, so that any effect it captures is attributable to mobility alone. The experiments cannot switch off this mobility, since every microgel grain is free to move. We isolate the mechanical role of obstacle mobility with a minimal \emph{active remodeler} particle: an active Brownian particle \cite{tenhagen-2011-jphys, romanczuk-2012-epjspectop}, of self-propulsion speed $v_{0}$ and rotational diffusivity $\mathcal{D}_{\mathrm{R}}$, at persistence P\'{e}clet number $\mathrm{Pe} = v_{0} / (a_{\mathrm{p}} \, \mathcal{D}_{\mathrm{R}})$ = 30 --- matched to the free-swimming value measured for the cells in our experiments (figure \ref{fig:fig-1}) --- moving through a disordered field of rigid, non-deformable disks of radius $a_{\mathrm{obs}}$ (= $a_{\mathrm{p}}$, i.e. $\alpha$ = 1, the same order-unity regime as the experiments) at obstacle area fractions $\psi$ --- where $\psi$ denotes the two-dimensional area fraction of the simulations, as distinct from the experimental three-dimensional effective volume fraction $\phi$ (see Methods for governing equations and numerical solution methodology). \\

Transport of active particles through fixed disordered or porous landscapes is by now well characterized: rigid obstacles reduce the effective diffusivity and, above a critical density, can localize the swimmer altogether --- a picture developed for active Brownian particles in random obstacle arrays \cite{chepizhko-2013-prl, reichhardt-2014-pre, zeitz-2017-epje} and realized experimentally for bacterial hopping and trapping in porous media \cite{bhattacharjee-2019-natcommun, bhattacharjee-2019-sm, creppy-2019-prf}, with matching theory for run-and-tumble dispersion through disordered media \cite{saintillan-2023-pre}. Far less explored is the case where the obstacles can be displaced by the swimmer --- the situation of a swimmer in a soft, remodelable matrix --- which on the lattice (i.e. for a passive particle) has only recently been shown to change the transport metrics qualitatively: replacing localization by slow, sustained spreading when a walker can push clusters of obstacle aside \cite{bonomo-2023-prr, singh-2026-pre, bonomo-2026-prl, shitrit-2026-arxiv}. Our model places this mechanism in an active, off-lattice, experimentally-anchored setting. \\

Each obstacle is a rigid (non-deformable) disk held against interpenetration by a stiff contact but free to translate against its drag $\gamma_{\mathrm{obs}}$; it is displaced, not deformed. The mobility ratio $\Gamma = \gamma_{\mathrm{obs}} / \gamma_{\mathrm{p}}$ ($\gamma_{\mathrm{p}}$ being the drag experienced by the cell) sets how readily an obstacle moves relative to the swimmer, from $\Gamma \rightarrow \infty$ (fixed; figures \ref{fig:fig-4}A.i and \ref{fig:fig-4}B, and movie SM3) to $\Gamma \ll$ 1 (freely displaced; figures \ref{fig:fig-4}A.iii and \ref{fig:fig-4}B, and movie SM3, for $\Gamma$ = 0.03), with $\Gamma$ = 0.3 (figures \ref{fig:fig-4}A.ii and \ref{fig:fig-4}B, and movie SM3) as a representative intermediate mobility. The swimmer exerts at most a finite stall force $f_{\mathrm{stall}} = \gamma_{\mathrm{p}} \, v_{0}$ --- of the order of the flagellar forces measured for \emph{C. reinhardtii} \cite{bayly-2011-biophysj, boddeker-2020-jrsi} --- so it cannot pass through an obstacle it is unable to move; it must go around it or push it aside. Crucially, this force cap applies only to what the swimmer \emph{delivers} to an obstacle. An immovable  ($\Gamma \rightarrow \infty$) obstacle's reaction force on the swimmer is uncapped, so it can push back with whatever force is needed to stop the swimmer, and blocks it outright rather than eventually getting displaced once enough force builds up (see Methods and figure \ref{fig:fig-s6}). This is also why mobility is set here by the drag ratio $\Gamma = \gamma_{\mathrm{obs}} / \gamma_{\mathrm{p}}$ rather than by the size ratio $\alpha = a_{\mathrm{obs}} / a_{\mathrm{p}}$ directly: $a_{\mathrm{obs}}$ and $a_{\mathrm{p}}$ govern the dynamical response to that contact, i.e. how far each body actually moves for a given force. It is this force-to-velocity relationship, set by drag, that determines whether a pushed obstacle displaces or holds firm, independent of its size. Both $a_{\mathrm{obs}}$ and $a_{\mathrm{p}}$ are held fixed throughout the numerical simulations presented here, so $\Gamma$ is the only parameter varied, and any difference between two values of $\Gamma$ isolates the effect of mobility alone. \\

We quantify transport by the normalized exploration $\chi = \mathcal{D}_{\mathrm{eff}} (\psi, \Gamma) / \mathcal{D}_{\mathrm{free}}$, the long-time, ensemble-averaged diffusivity scaled by its obstacle-free value --- fitted from the same late-time MSD slope as in the experiments, over the wall-guarded window ($\chi$ = 1 free, $\chi \rightarrow$ 0 arrested; see Methods and figure \ref{fig:fig-s6}B for further details). The central result is summarized in figure \ref{fig:fig-4}C: the normalized exploration $\chi$ as a function of $\psi$, shown for a representative set of $\Gamma$ values spanning the fixed limit to the most mobile cases simulated (see figure \ref{fig:fig-s6}C for a detailed phase map of $\chi (\psi, \Gamma)$). At every obstacle density, more mobile obstacles preserve more exploration. In the fixed limit, $\chi$ falls steeply and monotonically with crowding, and reaches complete arrest by $\psi \approx$ 0.25. Fixed obstacles therefore do not merely slow the swimmer; above a threshold density they localize it. This is the continuum-active counterpart of the localization transition of the Lorentz model \cite{hofling-2013-rpp} and of the arrest reported for active Brownian particles in fixed obstacle arrays \cite{chepizhko-2013-prl, zeitz-2017-epje}. Obstacle mobility postpones this collapse dramatically: at high mobility ($\Gamma$ = 0.03), $\chi$ remains substantially above its fixed-limit value at every $\psi$ studied, and at a moderate mobility ($\Gamma$ = 0.3), $\chi$ likewise stays well above the fixed branch across the density range --- the fixed limit is the lower envelope of the entire family of curves, so no amount of crowding makes a mobile medium less explorable than a fixed one at the same density.  \\

The contact-force records (figure \ref{fig:fig-s6}A) show why: the force the swimmer delivers to the medium --- bounded per contact by the stall force (see Methods) --- is substantially higher against fixed obstacles, where each grain moves aside after only a brief contact. The reaction the swimmer feels is by contrast uncapped for fixed obstacles (see Methods), so a fixed grain blocks it absolutely rather than being pushed through. That mobility converts a localized swimmer into a freely spreading one is the continuum, active analog of the escape mechanism established for the pushy random walk on a lattice \cite{bonomo-2026-prl}, in which allowing a tracer to displace clusters of obstacles replaces localization with slow, sustained spreading; it stands in deliberate contrast to the classical fixed-obstacle picture \cite{bhattacharjee-2019-natcommun, hofling-2013-rpp}, which our fixed limit reproduces. \\ 

Mobility is not, however, unlimited in what it can rescue. Beyond a density approaching the maximum coverage of equal disks ($\psi \approx$ 0.78) --- below the random close-packing ceiling for hard disks \cite{donev-2004-japplphys, donev-2005-pre} ($\psi \approx$ 0.82--0.84) and well below hexagonal close packing \cite{book-chaikin} ($\psi \approx$ 0.9069) --- all mobility ratios collapse together to $\chi \rightarrow$ 0, including the most freely displaced case (figure \ref{fig:fig-4}C): a swimmer of bounded propulsive force can no longer displace grains that are mutually jammed against their neighbors before either limit is reached. The relief that mobility provides therefore has a ceiling, set by the rigidity of the surrounding packing rather than the swimmer. Mechanically, this is the jamming limit of the obstacle network itself \cite{cates-1998-prl, liu-1998-nature, ohern-2003-pre}: once the grains bear load through a system-spanning contact network, displacing any one of them requires rearranging many. The same threshold logic governs a driven probe in a colloidal glass, which delocalizes only when the applied force exceeds the yield resistance of its neighbors \cite{habdas-2004-epl, gazuz-2009-prl}. Here, that criterion is the remodeling ratio $\mathcal{R} = f_{\mathrm{stall}} / f_{\mathrm{cage}} \geq$ 1, $f_{\mathrm{cage}}$ being the force required to displace a single grain from its cage in the jammed packing. Both forces entering $\mathcal{R}$ can be estimated from measurements in our system. Direct micropipette measurements give a flagellar force of 26 $\pm$ 5 \SI{}{\text{\pico\newton}} for \emph{C. reinhardtii} \cite{boddeker-2020-jrsi}. The yield stress resolved at our highest concentration ($\sigma_{\mathrm{y}} \approx$ 0.03 \SI{}{\text{\pascal}}), together with a grain scale $a_{\mathrm{obs}} \approx$ 10 \SI{}{\text{\micro\meter}}, gives a cage force $f_{\mathrm{cage}} \sim \sigma_{\mathrm{y}} \, a_{\mathrm{obs}}^{2} \approx$ 3 \SI{}{\text{\pico\newton}}. The remodeling ratio in the experiments is therefore $\mathcal{R} \approx$ 9, and remains above unity across the plausible range of grain sizes, 5--20 \SI{}{\text{\micro\meter}} (figure \ref{fig:fig-3}E). The cell's propulsive force therefore exceeds the force needed to displace a single grain by close to an order of magnitude, even in the most crowded sample we studied: remodeling is thus expected on force grounds and not merely observed. This also locates our experiments firmly on the mobile side of the model's operating boundary, consistent with the persistence of remodeling in figures \ref{fig:fig-3}A.i and \ref{fig:fig-3}A.ii (and movie SM2) at every $\phi$. \\

The model's own hop-and-trap statistics (figure \ref{fig:fig-4}D, computed identically to the experimental pipeline) let us ask, at the level of trends rather than matched values, whether the experimental system's confinement-driven kinematics track the model's mobile or fixed scenario. The mobile ($\Gamma$ = 0.3; circular markers with left and bottom axes in figure \ref{fig:fig-4}D) series's trap time RMST (restricted mean survival time; see Methods) declines substantially and monotonically with $\psi$, while the fixed (square markers with left and bottom axes in figure \ref{fig:fig-4}D) series's RMST --- over the range where it remains defined --- stays comparatively flat rather than declining. So, the two series diverge from each other. The experimental RMST (pentagram markers with top and right axes in figure \ref{fig:fig-4}D) likewise declines with increasing crowding, tracking the mobile series's direction rather than the fixed series. We emphasize that $\psi$ and $\phi$ are not the same physical quantity (see Methods): so this is a comparison of trend, not of matched values. Additionally, the fixed series's RMST in figure \ref{fig:fig-4}D is deliberately shown only over the $\psi$ range where it is defined at all, and this range is narrow. RMST is computed only from realizations that are not permanently arrested by the end of the observation window. Once every realization at a given $\psi$ is arrested, RMST is simply undefined at that $\psi$. For the fixed condition this happens over a comparatively low $\psi$, leaving only a handful of $\psi$ values at which RMST can be reported at all. The mobile condition never saturates this way over the range studied. \\
 
Thus, our numerical simulations isolate a variable the experiments could not: holding density, persistence, and obstacle size fixed, making the obstacles mobile --- without any change in their shape --- is sufficient to convert a medium that permanently cages the swimmer into one it explores freely. Since the grains in the experiments are also deformed as cells pass (figures \ref{fig:fig-3}A.i and \ref{fig:fig-3}A.ii, and movie SM2), this identifies mobility as a \emph{sufficient} mechanism, but not the only one at work in the experiment (see figure \ref{fig:fig-s9} for a discussion on grain compliance). Our simulations provide a natural mechanistic reading of the experiments --- the simulated obstacle displacement is the counterpart of the microgel remodeling imaged in figures \ref{fig:fig-3}A.i and \ref{fig:fig-3}A.ii (also movie SM2) --- and a candidate explanation for why cells are able to explore the densest gels rather than arresting within them, though the model is minimal and we do not attempt to reproduce the experiments quantitatively. 
 
\section{Conclusions and outlook}

We have tracked the motile uniceullar microalga \emph{C. reinhardtii} through a soft, disordered obstacle field, with a grain size matching the swimmer's body size, to understand how microscopic confinement shapes swimming behavior. Rather than acting as a static barrier, the soft medium interacts dynamically with the alga. Crowding reduces exploration and collapses the gap between the alga's two swimming gaits, much like a rigid barrier would. However, run lengths decrease and hops shorten without longer trapping times --- revealing a dynamic medium whose resistance adapts in real to the swimming microorganism. \\

We find that the cells visibly displace and rearrange the microgel as they pass, at every packing fraction studied. The confinement a cell experiences is therefore set in part by the cell itself: it does not wait for a fluctuation to free it from a cage, because it makes its own path. Thus, \emph{C. reinhardtii} in the complex environment studied here is in neither the continuum viscoplastic limit of a swimmer much larger than the microstructure ($\alpha \ll$ 1), nor the pore-threading limit of bacteria much smaller than the microstructure ($\alpha \gg$ 1). Instead, here the the discrete grain-specific length scale and the swimmer body size are comparable (i.e. $\alpha \approx$ 1), and the cell is able to reconfigure the individual soft elements comprising the medium. \\

Since deformability of the medium cannot be switched off in the experiments, we construct a numerical model featuring an active Brownian particle exploring a disordered medium comprised of obstacles of the same size as the active particle. Obstacle mobility becomes a control parameter, and our simulations show that mobility alone is sufficient to shift the area fraction of obstacles where permanent confinement sets in by roughly a factor of three as compared to immobile obstacles. This is the continuum, active counterpart of the lattice ``pushy'' random walk, in which a walker able to displace obstacles escapes localization altogether and spreads indefinitely instead \cite{bonomo-2023-prr, bonomo-2026-prl}. The relief from trapping is not unbounded: it has a ceiling, set by the jamming of the obstacle network itself against a swimmer of finite propulsive force. Hence, the model reproduces the threshold structure (or ``yield stress'') familiar from driven probes in dense amorphous media, where escape requires only that the driving forces exceed the local resistance of the surroundings \cite{habdas-2004-epl, gazuz-2009-prl}. The difference here is that the probe is self-propelled and the resistance is tunable through obstacle mobility rather than through the driving force. The model thus offers obstacle mobility as an independent control parameter for transport in soft, disordered matter, and a candidate mechanistic reading of why cells keep moving through the densest gels instead of becoming permanently arrested in them. \\

This work has a natural scope, which bounds how far these conclusions extend and sets the agenda for what comes next. The active-remodeler model proposed here is deliberately minimal --- a single active Brownian particle among soft disks --- and reproduces the experimental trends qualitatively. However, the model omits recovery of the displaced obstacles toward their original positions once the swimmer has passed --- obstacles in the model stay wherever they are pushed --- as well as the fully three-dimensional geometry of both swimmer and medium, grain polydispersity, hydrodynamics, and the finite geometry and beating of the flagella.  \\

These open ends point to concrete next steps. A companion numerical study can pursue the statistical physics the present minimal model only gestures at --- separating grain compliance from grain mobility as independent axes (a control the present rigid-obstacle model deliberately foregoes; see figure \ref{fig:fig-s9} and movie SM4 for a preliminary treatment showing that grain compliance alone also partially relieves suppression) --- connecting this active, off-lattice system rigorously to the pushy-random-walk universality \cite{bonomo-2026-prl}. Resolving the elastic component of the supernatant, and measuring motility in a grain-free polymer solution at matched low-shear viscosity, would separate the viscous and elastic contributions of the enhancement and reconcile the sign with the hindrance reported for this alga in dilute elastic fluids \cite{teran-2010-prl, thomases-2014-prl, qin-2015-scirep}. More broadly, applying the same measurements across swimmer sizes and gaits would test whether obstacle mobility is a general axis of microbial transport in soft media. A first step in this direction is already in hand: the mutant strains used here to vary the remodeling ratio $\mathcal{R}$ through the swimmer (figures \ref{fig:fig-s7} and \ref{fig:fig-s8}) already show that flagellar architecture has an influence on how far the medium can affect the swimmer's motion. If obstacle mobility does turn out to be a general axis, it reframes the microbial confinement problem for the natural habitats that motivated this study --- soil pore networks, mucus, and tissue matrices --- as one not of navigating a fixed landscape, but of reshaping a responsive one, and offers the same reframing for the design of microrobots meant to move through soft media. 
  
\section{Methods}

\subsection{\emph{Chlamydomonas reinhardtii} culture}

\emph{Chlamydomonas reinhardtii} (strain 11/32A, referred to as ``wild type (WT)''; purchased from Culture Collection of Algae and Protozoa, Scotland, United Kingdom) was grown axenically in Tris-acetate-phosphate (``TAP''; Gibco, Fisher Scientific) growth medium \cite{catalan-2025-natprotoc} on a 16:8 \SI{}{\text{\hour}} light-dark cycle at 21\SI{}{\text{\celsius}}. Cells were harvested during day 5 (exponential phase) of their life cycle \cite{catalan-2025-natprotoc}, and diluted to a density of $\approx$ 1.42 $\times$ 10$^{\text{5}}$ \SI{}{\text{cells.\milli\liter\tothe{-1}}} for each experiment by adding fresh TAP medium. Mutant strains 2894, with asynchronous flagella, and 1926, with a single flagellum, (both purchased from Chlamydomonas Resource Center, University of Minnesota, United States), used for the strain comparisons in figures \ref{fig:fig-s7} and \ref{fig:fig-s8}, were cultured identically. 

\subsection{Microgel dispersion preparation}

Soft, heterogeneous confinement was imposed with Carbopol (Ultrez 10; Lubrizol) microgel suspensions prepared in TAP medium. A 1\% (by weight) stock was prepared by dispersing dry Carbopol powder in TAP and gently stirring overnight to achieve a homogeneous dispersion. Triethanolamine (Sigma-Aldrich) was then added dropwise under stirring until a pH of 7 was achieved, thus inducing swelling and gelation. The stock equilibrated overnight and was diluted to the required weight fractions for the experiments by adding the appropriate mass of TAP medium. Diluted gels were rested for at least two hours before any experimental measurement. For fluorescence microscopy of the microgel suspension, the cationic dye Rhodamine 6G (Sigma-Aldrich) was added in small volume (10 \SI{}{\text{\micro M}} stock) during dispersion; it binds electrostatically to the anionic microgel, selectively labeling the polymer phase \cite{dinkgreve-2018-jor}. 

\subsection{Effective volume fraction}

Since the swollen microgel packing, not the dry polymer mass, sets the crowding a cell experiences, we converted the weight fraction to an effective volume fraction $\phi$ by inverting the Krieger-Dougherty relation \cite{krieger-1959-transsocrheol}
\begin{align}
    \frac{\eta_{0}}{\eta_{\mathrm{s}}} = \left( 1 - \frac{\phi}{\phi_{\max}} \right)^{ - [\eta] \, \phi_{\max}},
    \label{eq:KD-fit}
\end{align}
where $\eta_{0}$ is the low-shear viscosity of the microgel dispersion (evaluated here at a shear rate of 10 \SI{}{\text{\second\tothe{-1}}}), $\eta_{\mathrm{s}}$ the viscosity of the solvent phase (TAP medium), $\phi_{\max}$ (= 0.64) the maximum packing fraction, and $[\eta]$ (= 2.5) the intrinsic viscosity; $\eta_{0}$ for the different Carbopol weight fractions and $\eta_{\mathrm{s}}$ were obtained from steady-shear rheometry (figure \ref{fig:fig-s1}A). This inversion maps Carbopol weight fractions to effective volume fractions $\phi$ (figure \ref{fig:fig-s1}B). Since the most concentrated samples are already weakly shear thinning at this rate (figure \ref{fig:fig-s1}A), this yields a conservative (lower) estimate of $\phi$. A sensitivity sweep over $[\eta]$ and $\phi_{\max}$ left the qualitative progression unchanged. 

\subsection{Microscopy and cell tracking}

Cells were loaded into a rectangular (6 \SI{}{\text{\milli\meter}} $\times$ 8 \SI{}{\text{\milli\meter}}) quasi-two-dimensional chamber of height 0.22 \SI{}{\text{\milli\meter}}, fabricated using two microscope glass cover slips (VWR) and double-sided adhesive tape. The cell-loaded chamber was imaged in a single focal plane by brightfield microscopy on an inverted microscope (Ti-E; Nikon) equipped with a 5$\times$ objective (CFI TU Plan FLUOR, 0.15 NA; Nikon) at a frame rate of 10 \SI{}{\text{\hertz}} (Prime BSI; Teledyne Photometrics), with a spatial calibration of 1.02 \SI{}{\text{\micro\meter.pixel\tothe{-1}}}. To avoid phototactic bias of the cellular motility \cite{hegemann-2008-arplantbiol, kreis-2018-natphys, catalan-2025-natprotoc}, all recordings were made under a continuous 5 \SI{}{\text{\watt.\meter\tothe{-2}}} illumination (MXA20434 tungsten-halogen lamp; Nikon) passed through a 665 \SI{}{\text{\nano\meter}} long-pass filter (Thorlabs). \\

Trajectories were extracted using a Trackpy-based \cite{web-trackpy} custom Python-script using the Crocker-Grier algorithm \cite{crocker-1996-jcis}, linked across frames, and exported as time series of centroid positions and instantaneous speed. At least 500 trajectories were analyzed per $\phi$-condition, pooled from at least three independent sample preparations. Since imaging is confined to one plane, out-of-plane excursions of a three-dimensional swimmer such as \emph{C. reinhardtii} bias the in-plane speed downward; we restricted our analysis to tracks remaining in focus for at least 3 \SI{}{\text{\second}}. Three-dimensional particle tracking at representative $\phi$-conditions, performed on a confocal microscope (Nikon Ti2 with a Nikon C2 point-scanning confocal and a Nikon CFI Plan Apo 10$\times$ NA 0.45 objective) with the $z$-stack functionality (13 slices with step size of 5 \SI{}{\text{\micro\meter}}), further confirms the primarily two-dimensional nature of the trajectories in the experiments (see figure \ref{fig:fig-s2}). 

\subsection{Mean-squared displacement and effective diffusivity}

For each $\phi$-condition we computed the ensemble-averaged MSD $\langle \vert \Delta \mathbf{r} (\tau) \vert^{2} \rangle$ over lag times $\tau$ upto half the maximum track duration, avoiding undersampled long-time correlations. The ballistic-to-diffusive crossover time $\tau_{\mathrm{c}}$ and long-time effective diffusivity $\mathcal{D}_{\mathrm{eff}}$ were obtained by fitting the active Brownian particle form \cite{tenhagen-2011-jphys}
\begin{align}
    \langle \vert \Delta \mathbf{r} (\tau) \vert^{2} \rangle = 4 \, \mathcal{D}_{\mathrm{eff}} \left( \tau - \tau_{\mathrm{c}} \left( 1 - \exp (- \tau / \tau_{\mathrm{c}}) \right) \right) .
\end{align}
The above expression reduces to familiar behavior in two limits. For $\tau \ll \tau_{\mathrm{c}}$, expanding the exponential gives $\langle \vert \Delta \mathbf{r} (\tau) \vert^{2} \rangle \approx \left(2 \, \mathcal{D}_{\mathrm{eff}} / \tau_{\mathrm{c}} \right) \tau^{2}$ --- ballistic motion, at an effective mean-squared speed $v^{2} = 2 \, \mathcal{D}_{\mathrm{eff}} / \tau_{\mathrm{c}}$. For $\tau \gg \tau_{\mathrm{c}}$, the exponential term vanishes and $\langle \vert \Delta \mathbf{r} (\tau) \vert^{2} \rangle \approx 4 \, \mathcal{D}_{\mathrm{eff}} \left( \tau - \tau_{\mathrm{c}} \right)$ --- purely diffusive, with $\mathcal{D}_{\mathrm{eff}}$ the asymptotic long-time diffusivity and $\tau_{\mathrm{c}}$ contributing only a constant time offset. $\mathcal{D}_{\mathrm{eff}}$, $\tau_{\mathrm{c}}$, and $v$ are therefore not three independent quantities but two: once $\mathcal{D}_{\mathrm{eff}}$ and $\tau_{\mathrm{c}}$ are fitted from the data, $v$ is fixed via $v = \sqrt{2 \, \mathcal{D}_{\mathrm{eff}} / \tau_{\mathrm{c}}}$. 

\subsection{Speed distributions and bimodality}

Instantaneous speed distributions (figure \ref{fig:fig-2}A) pool all in-focus frames per $\phi$-condition ($\approx$ 10$^{\text{6}}$ speeds). Since a 0.1 \SI{}{\text{\micro\meter}} displacement at 10 \SI{}{\text{\hertz}} corresponds to 1 \SI{}{\text{\micro\meter.\second\tothe{-1}}}, near the localization limit, and $\approx$ 13\% of frames in TAP medium ($\phi$ = 0) fall below this value, speeds below $v_{\min}$ = 2 \SI{}{\text{\micro\meter.\second\tothe{-1}}} were excluded from further analysis. Multimodality of the instantaneous speed distributions was tested with Hartigan's dip test \cite{hartigan-1985-annstat}. Run and tumble sub-states were separated by modeling the pooled speed distribution at each $\phi$ as a weighted sum of two normal distributions --- a low-speed component (tumbles) and a high-speed component (runs) --- fitted by expectation-maximization \cite{book-mclachlan, dempster-1977-jrstatsocb}. Each frame is then assigned to whichever of the two components has the higher posterior probability, giving a per-frame run/tumble label from which state mean speeds and the run-time fraction are computed directly; both were bootstrapped (200 resamples) for confidence intervals. Robustness of the collapse to the $v_{\min}$ choice was confirmed by sweeping over $v_{\min}$ (figure \ref{fig:fig-s3}A). 

\subsection{Velocity autocorrelation and persistence quantification}

Directional memory was quantified from the velocity autocorrelation function \cite{book-furst}
\begin{align}
    C (\tau) = \langle \mathbf{v} (t) \cdot \mathbf{v} (t + \tau) \rangle ,
\end{align}
and its normalized form $\hat{C} (\tau) = C (\tau) / C (0)$, computed per track over lag times $\tau$ up to half the track duration, and fitted to $\hat{C} (\tau) = \exp (- \tau / \tau_{\mathrm{p}})$ to obtain the persistence time $\tau_{\mathrm{p}}$. The persistence length is the derived quantity $L_{\mathrm{p}} = \tau_{\mathrm{p}} \, v_{\mathrm{m}}$, with $v_{\mathrm{m}}$ the mean track speed. We verified $L_{\mathrm{p}} = \tau_{\mathrm{p}} \, v_{\mathrm{m}}$ to machine precision, so $L_{\mathrm{p}}$ is not an independent observable; its spread compounds those of $\tau_{\mathrm{p}}$ and $v_{\mathrm{m}}$. Because $\tau_{\mathrm{p}}$ and $L_{\mathrm{p}}$ are strictly positive and right-skewed, their kernel-density estimates were computed in log-space, which avoids the artificial boundary bias a density estimator otherwise produces near zero for positive-only data, and mapped back to a linear-scale density using the Jacobian of the log transform. 

\subsection{Hop-and-trap segmentation}

Trajectories were segmented into alternating hops and traps to compare with the hopping-and-trapping framework for rigid porous media \cite{bhattacharjee-2019-natcommun, sinaasappel-2025-prl}. To avoid biasing the segmentation by the $\phi$-dependent speed distribution, the threshold was set relative to each $\phi$-condition's own median speed (0.5 $\times$ median); a binary above/below-threshold series was median-filtered to remove single-frame flicker, and contiguous runs defined hops (displacement between endpoints) and traps (duration). Three controls were applied (figure \ref{fig:fig-s4}): (i) tracks with median speed $<$ 2 \SI{}{\text{\micro\meter.\second\tothe{-1}}} were excluded as immotile (their fraction falls from 14.8\% at $\phi$ = 0 to 0\% at the highest $\phi$ tested; consistent with cells settling out of the focal plane at $\phi$ = 0 over the course of a long recording, but held in suspension by the microgel dispersion at higher $\phi$); (ii) the analysis was repeated for fixed thresholds of 3, 5, and 8 \SI{}{\text{\micro\meter.\second\tothe{-1}}}, showing that sweeping a single fixed threshold across a $\phi$-dependent speed distribution produces spurious trends in the resulting hop and trap statistics; and (iii) since median track duration is non-monotonic in $\phi$, all $\phi$-conditions were restricted to a common 10.0 \SI{}{\text{\second}} observation window. Many trapping episodes are still ongoing when a track ends: for these, the true trap duration is known only to exceed some lower bound, not measured exactly --- a ``censored'' observation, in survival-analysis terms. Treating a censored trap as though it ended the instant the track did would systematically underestimate trap duration, especially at the packing fractions or strains where censoring is common. Trapping-time ($\tau_{\mathrm{trap}}$) distributions were estimated by the Kaplan-Meier method \cite{kaplan-1958-jasa}, which incorporates each censored observation as the lower bound it actually is rather than discarding it or treating it as a completed measurement. The median trap time is read as the time at which the resulting survival function $S (\tau_{\mathrm{trap}})$ crosses 0.5. No functional form was fitted to the survival curves. The restricted mean survival time (RMST) was computed from the area under the $S (\tau_{\mathrm{trap}})$ curves up to a fixed horizon. \\

To benchmark the trapping-time result, each $\phi$ = 0 trajectory was re-interpolated onto a time axis stretched by $1 / \lambda (\phi)$, where $\lambda (\phi)$ is the measured ratio of median speed at that $\phi$ to the median speed at $\phi$ = 0 for that strain, and re-sampled at the original frame interval. This keeps the path and the sequence of fast and slow episodes unchanged, altering only how quickly they are executed --- exactly what a medium that slows down the swimmer without being deformed by it would do. The dilated tracks were passed through an identical cleaning, segmentation, and survival pipeline (figure \ref{fig:fig-s8}).

\subsection{Confocal imaging of medium remodeling}

The Rhodamine-6G-labeled microgel dispersion \cite{dinkgreve-2018-jor} was imaged by dual-channel confocal microscopy (Nikon Ti2 inverted microscope equipped with a Nikon C2 point-scanning confocal and a Nikon CFI Plan Apo 10$\times$ NA 0.45 objective, 20 \SI{}{\text{\micro\meter}} pinhole, 525/50 and 593/40 + 650 LP emission filters) while cells swam through it, at 1.034 \SI{}{\text{\hertz}} with a spatial resolution of 1.657 \SI{}{\text{\micro\meter.pixel\tothe{-1}}} (DU3 photomultiplier unit; Nikon), capturing the deformation and elastic recovery of the microgel around individual swimmers. \\

The characteristic pore size of the labeled microgel dispersion was quantified by granulometry \cite{book-matheron, book-serra, book-soille}. Confocal images were thresholded into microgel and void phases; the void phase was then subjected to morphological opening with a disk-shaped structuring element of increasing diameter $D$. The opened area $A(D)$ is the union of all disks of diameter $D$ that fit within the void, so $A(D) / A(0)$ is the cumulative fraction of void belonging to pores of at least that size, $- \mathrm{d} (A(D) / A(0)) / \mathrm{d} D$ is the pore-size density. \\

To render the confocal remodeling micrographs as a quantitative map of gel displacement rather than fluorescence intensity, each frame pair in the labeled-microgel channel was passed through dense Farneback optical flow \cite{farneback-conf}, giving a per-pixel displacement field between consecutive frames. The raw flow field was smoothed with a small Gaussian kernel to suppress pixel-scale noise before display, and pixels outside the labeled-gel signal (identified by a minimum-intensity mask on the red channel) were excluded, so that the colormap shows displacement only where the gel is actually present rather than extrapolating into the background. The same color scale is shared across all frames and both $\phi$-conditions shown in figure \ref{fig:fig-3}. The cell itself is rendered as a fixed-color mask over its own segemented outline in the green channel.

\subsection{Numerical model}

\subsubsection{Governing equations}

We modeled the swimmer as a single active Brownian particle \cite{romanczuk-2012-epjspectop, tenhagen-2011-jphys, bechinger-2016-rmp} of radius $a_{\mathrm{p}}$ moving in two dimensions through a disordered field of rigid (non-deformable), repulsive, circular obstacles of radius $a_{\mathrm{obs}} = a_{\mathrm{p}}$ that translate but do not change shape, in a square box of side $L = 320 \, a_{\mathrm{p}}$ with reflecting walls. The particle self-propels at a fixed speed $v_{0}$ along its orientation $\mathbf{e}_{\theta} = (\cos \theta, \sin \theta)$, which undergoes rotational diffusion with coefficient $\mathcal{D}_{\mathrm{R}}$. Writing the particle position as $\mathbf{r} (t)$ and obstacle centers as $\{ \mathbf{c}_{i} (t) \}$, the overdamped equations of motion of the particle are
\begin{subequations}
    \begin{align}
        \gamma_{\mathrm{p}} \frac{\mathrm{d} \mathbf{r}}{\mathrm{d} t} &= f_{\mathrm{stall}} \, \mathbf{e}_{\theta} + \sum_{i} \mathbf{F}_{\mathrm{p} i} , \\
        \frac{\mathrm{d} \theta}{\mathrm{d} t} &= \sqrt{2 \, \mathcal{D}_{\mathrm{R}}} \, \xi (t) ,
    \end{align}
\end{subequations}
where $\gamma_{\mathrm{p}}$ is the particle drag, $\xi (t)$ Gaussian white noise with $\langle \xi (t) \rangle = 0$ and $\langle \xi (t) \, \xi (t^{\prime}) \rangle = \delta (t - t^{\prime})$ ($\delta$ denoting the Dirac delta function), and $\mathbf{F}_{\mathrm{p} i}$ the harmonic contact force exerted on the particle by obstacle $i$. Defining the center-to-center separation $\mathbf{x}_{\mathrm{p} i} = \mathbf{r} - \mathbf{c}_{i}$, with magnitude $x_{\mathrm{p} i} = \vert \mathbf{x}_{\mathrm{p} i} \vert$ and unit vector $\mathbf{e}_{\mathrm{p} i} = \mathbf{x}_{\mathrm{p} i} / x_{\mathrm{p} i}$, the overlap is $\zeta_{\mathrm{p} i} = (a_{\mathrm{p}} + a_{\mathrm{obs}}) - x_{\mathrm{p} i}$, and
\begin{align}
    \mathbf{F}_{\mathrm{p} i} = \mathcal{K} \, \zeta_{\mathrm{p} i} \, \mathbf{e}_{\mathrm{p} i} \, \Theta (\zeta_{\mathrm{p} i}) ,
\end{align}
with stiffness $\mathcal{K}$ and $\Theta$ the Heaviside step function: the force is purely repulsive and acts only in contact ($\zeta_{\mathrm{p} i} > 0$). The propulsive drive is capped at the stall value $f_{\mathrm{stall}} = \gamma_{\mathrm{p}} \, v_{0}$: the swimmer exerts at most this force on the medium, so it cannot displace an obstacle it is unable to move. The cap applies only to the force the swimmer delivers to a mobile obstacle; the reaction that a fixed (immovable) obstacle exerts back on the swimmer is uncapped, so the swimmer is genuinely blocked by obstacles it cannot displace rather than passing through them. \\

Each obstacle is itself mobile and overdamped, evolving under contact forces from its neighbors and from the swimmer,
\begin{align}
    \gamma_{\mathrm{obs}} \frac{\mathrm{d} \mathbf{c}_{i}}{\mathrm{d} t} = \sum_{j \neq i} \mathbf{F}_{ij} - \bar{\mathbf{F}}_{\mathrm{p} i} ,
\end{align}
where
\begin{align}
    \mathbf{F}_{ij} = \mathcal{K} \, \zeta_{ij} \, \mathbf{e}_{ij} \, \Theta (\zeta_{ij}) ,
    \label{eq:force-obstacle}
\end{align}
is the obstacle-obstacle contact force with $\mathbf{x}_{ij} = \mathbf{c}_{i} - \mathbf{c}_{j}$ and overlap $\zeta_{ij} = 2 \, a_{\mathrm{obs}} - x_{ij}$, $\gamma_{\mathrm{obs}}$ the obstacle drag, and
\begin{align}
    \bar{\mathbf{F}}_{\mathrm{p} i} = \min (\mathcal{K} \, \zeta_{\mathrm{p} i}, f_{\mathrm{stall}}) \, \mathbf{e}_{\mathrm{p} i} \, \Theta (\zeta_{\mathrm{p} i})
\end{align}
is the stall-capped force the swimmer transmits to obstacle $i$. The single dimensionless group controlling obstacle mobility is $\Gamma = \gamma_{\mathrm{obs}} / \gamma_{\mathrm{p}}$: $\Gamma \rightarrow \infty$ pins the obstacles (fixed limit), while finite $\Gamma$ lets the swimmer displace them. The obstacles are rigid (non-deformable) throughout; the model represents grain rearrangement but not grain compliance (see figure \ref{fig:fig-s9} and movie SM4 for a preliminary treatment of grain compliance). The persistence P\'{e}clet number is $\mathrm{Pe} = v_{0} / (a_{\mathrm{p}} \, \mathcal{D}_{\mathrm{R}})$. Obstacle configurations were generated by random sequential placement \cite{widom-1966-jchemphys} at target area fraction $\psi$. We denote the obstacle density in the model $\psi$ to distinguish it from the experimental packing fraction $\phi$: $\phi$ is an effective three-dimensional volume fraction inferred from the suspension viscosity through a Krieger-Dougherty inversion, whereas $\psi$ is the two-dimensional area fraction of obstacle disks in the simulation plane, imposed directly. The two are conceptually distinct crowding measures --- a 3D volume fraction and a 2D area fraction are not numerically interchangeable. We therefore make no attempt to match them point-by-point, and compare experiment and model only at the level of trends across their respective density ranges. Non-dimensionalizing lengths by $a_{\mathrm{p}}$, times by $1 / \mathcal{D}_{\mathrm{R}}$, and forces by $f_{\mathrm{stall}}$, where the non-dimensional contact stiffness $\kappa = \mathcal{K} \, a_{\mathrm{p}} / f_{\mathrm{stall}}$, the model is fully specified by $(\psi, \Gamma, \mathrm{Pe}, \kappa)$. \\

Two-dimensionality is a deliberate choice for three reasons. First, it is the minimal setting in which obstacle mobility can be isolated as a single control parameter and swept exhaustively over the full $(\psi, \Gamma)$ plane --- the same minimality argument used also to justify omitting grain compliance. Second, it keeps the model computationally tractable at the ensemble sizes (96 realizations per condition) needed for the exploration-trapping statistics in figure \ref{fig:fig-4}. Third, and most directly tied to the experiments themselves: the experimental comparison this model is built against is itself a planar analysis --- all recordings were acquired in a single focal plane (figure \ref{fig:fig-1}), and figure \ref{fig:fig-s2} shows directly that the out-of-plane contribution to cell displacement is subdominant to the in-plane contribution at every $\phi$ tested. A two-dimensional model is therefore not only the more tractable choice but also the dimensionally matched one: it targets the same planar projection of the dynamics that the experimental trajectories themselves resolve, rather than a full three-dimensional geometry that the experimental analysis could not be compared against on equal terms even if the model included it.

\subsubsection{Numerical solution}

The equations of motion were integrated with an Euler-Maruyama scheme \cite{book-kloeden, ermak-1978-jchemphys} at a time step of $0.01 / \mathcal{D}_{\mathrm{R}}$ for 36000 steps ($v_{0}$ = 1, $\mathcal{D}_{\mathrm{R}}$ = 1/30 so $\mathrm{Pe}$ = 30, $\mathcal{K}$ = 100, $f_{\mathrm{stall}}$ = 1, and $\gamma_{\mathrm{p}}$ = 1) using an in-house developed Python-based implementation \cite{web-zenodo}. Trajectories were sub-sampled at every $0.1 / \mathcal{D}_{\mathrm{R}}$ time step for analysis. For each $(\psi, \Gamma)$ we swept the obstacle area fraction $\psi$ over the range 0.05--0.90 and $\Gamma$ over 0.03--30, plus the fixed (immobile) obstacle limit, at 96 independent obstacle realizations per cell. Obstacle configurations were generated by random sequential placement \cite{widom-1966-jchemphys} at target area fraction $\psi$; where sequential placement stalled before reaching the target count --- as increasingly occurs above the $\psi \approx$ 0.55 jamming point of pure random sequential addition, and substantially so above $\psi \approx$ 0.78 --- the remaining obstacles were seeded at random and a brief pairwise relaxation reduced overlaps. Any residual overlap in this seed configuration (never exceeding $\approx$ 0.25 obstacle radii even at the highest $\psi$ studied) is resolved within the first few integration steps by the same obstacle-obstacle contact force (equation \eqref{eq:force-obstacle}) that governs the rest of the dynamics, and a direct check --- re-running the $\psi$ = 0.78 condition with a ten-fold longer relaxation pass --- changes the fitted long-time diffusivity by only 9\% and leaves the trapped fraction unchanged. Thus, obstacle disorder in the model arises entirely from this random placement process --- the model's off-lattice counterpart of the disordered microstructure that the experimental microgel dispersion itself acquires through unguided swelling and jamming. The active particle feels the full, uncapped elastic reaction force from each fixed (immobile) obstacle and cannot penetrate one it is unable to move; the stall cap $f_{\mathrm{stall}}$ limits only the force transmitted by the particle to a mobile obstacle (figure \ref{fig:fig-s6}A). Ensemble MSDs were computed as for the experiments, and the normalized diffusivity $\chi = \mathcal{D}_{\mathrm{eff}} / \mathcal{D}_{\mathrm{free}}$ was obtained by fitting the late-time slope over a time window chosen so that the particle's displacement stays below $L / 2$, excluding reflecting-wall saturation (figure \ref{fig:fig-s6}B); the free-particle diffusivity value $\mathcal{D}_{\mathrm{free}} = v_{0}^{2} / (2 \, \mathcal{D}_{\mathrm{R}}) = \mathrm{Pe} / 2$ sets the normalization. 

\section{Acknowledgments}

We acknowledge financial support from the Netherlands Organization for Scientific Research (NWO) through the OCENW.XS.24.1.139, OCENW.XS.25.2.188, and OCENW.XS.26.1.062 grants, and the European Molecular Biology Organization (EMBO) through grant number 11317. The authors acknowledge the use of the Anunna High Performance Computing facility of Wageningen University and Research for providing computational resources for this research. The authors sincerely thank Maziyar Jalaal for valuable discussions and experimental support, which were instrumental in shaping the early findings of this work. The authors are grateful for the technical assistance from Remco Fokkink and Raoul Fix during the fabrication of the experimental setup, and the Microspectroscopy Research Facility for assistance with confocal microscopy. The authors thank Jorge Arrieta, Quirine Braat, Eric Cl\'{e}ment, Sujit Datta, Wouter Ellenbroek, Silke Henkes, Liesbeth Janssen, Sander Kammeraat, Thomas Kodger, Anke Lindner, Arnold Mathijssen, Justin Tauber, Federico Toschi, and Idan Tuval for insightful discussions. 

\section{Author contributions}

A.B., N.S., and U.S. designed the experimental system. A.B. performed the experiments; A.B. and U.S. performed the experimental data analysis. U.S. developed the numerical model and performed the associated data analysis. All authors contributed to interpreting the results and writing the manuscript. 

\section{Competing interests}

The authors declare no competing interests.

\section{Data availability}

Data is available from the corresponding author upon request. 


\newpage

\setcounter{equation}{0}
\setcounter{figure}{0}
\setcounter{table}{0}

\renewcommand{\thefigure}{S\arabic{figure}}
\renewcommand{\thetable}{S\arabic{table}}
\renewcommand{\theequation}{S\arabic{equation}}

\section{Supplementary information}

\FloatBarrier

\subsection{Rheological characterization and volume-fraction assignment}

\begin{figure}
    \centering
    \includegraphics[width=\textwidth]{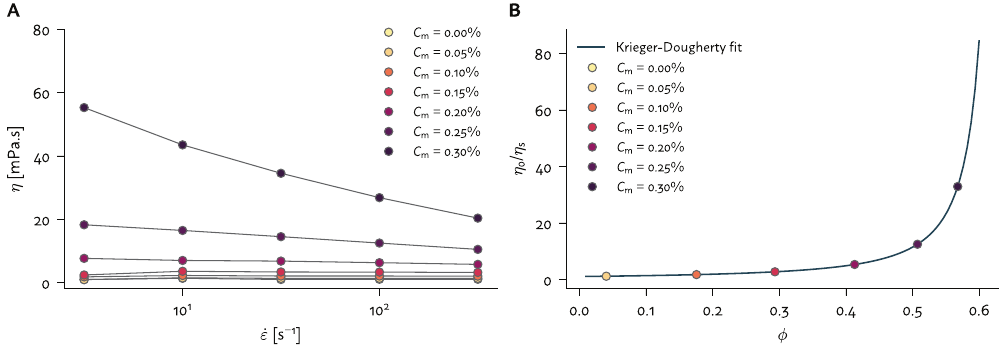}
    \caption{\textbf{Rheology and concentration-volume fraction mapping.} \textbf{A.} Steady-shear flow curves (shear viscosity $\eta$ versus shear rate $\dot{\varepsilon}$) for all Carbopol concentrations $C_{\mathrm{m}}$ (by weight). The suspensions range from effectively Newtonian at low $C_{\mathrm{m}}$ to weakly shear-thinning at high $C_{\mathrm{m}}$; no low-shear stress plateau is resolved. Yield stress and linear-viscoelastic moduli were at or below the rheometer noise floor except at $C_{\mathrm{m}}$ = 0.30\%, where $\sigma_{\mathrm{y}} \approx$ 0.03 \SI{}{\text{\pascal}} could be estimated. \textbf{B.} Relative low-shear viscosity $\eta_{0} / \eta_{\mathrm{s}}$, normalized by the viscosity of the TAP medium, versus grain volume fraction $\phi$, estimated from the Krieger-Dougherty relation (equation \eqref{eq:si-KD-fit}; solid line). Since $\phi$ is obtained by inverting equation \eqref{eq:si-KD-fit} from the measured $\eta_{0} / \eta_{\mathrm{s}}$, the data lie on the curve by construction. }
    \label{fig:fig-s1}
\end{figure}

Rheological measurements were performed on a stress-controlled rotational rheometer (MCR 501; Anton Paar GmbH) using a cone-and-plate geometry (1\SI{}{\text{\degree}} angle, 75 \SI{}{\text{\milli\meter}} diameter, and a mean gap of 0.1 \SI{}{\text{\milli\meter}}). All measurements were performed at 21\SI{}{\text{\celsius}} and in triplicate. \\

The experimental control parameter is the dry Carbopol mass concentration $C_{\mathrm{m}}$, but the physically relevant parameter is the volume fraction of swollen microgel grains. Flow curves for all Carbopol concentrations are shown in figure \ref{fig:fig-s1}A. The suspensions range from effectively Newtonian at low concentrations to weakly shear-thinning at the higher concentrations. The absolute viscosities are modest --- at most a few tens of times that of water --- and the shear thinning is mild, a factor of $\approx$ 3 over two decades of shear rate. \\

We attempted to characterize the yield stress ($\sigma_{\mathrm{y}}$) and the linear-viscoelastic moduli ($G^{\prime}$, $G^{\prime \prime}$) of these suspensions directly, but at the concentrations used here both quantities lie at or near the noise floor of our rheometer, and could not be determined reliably across the series. Only for the highest concentration ($C_{\mathrm{m}}$ = 0.30\%) could a yield stress be estimated, giving $\sigma_{\mathrm{y}} \approx$ 0.03 \SI{}{\text{\pascal}}. This value is itself instructive: it is more than three orders of magnitude below the yield stress of the concentrated Carbopol gels ($\sim$ 1--100 \SI{}{\text{\pascal}}) more commonly studied as model yield-stress fluids \cite{dinkgreve-2018-jor}, and confirms that our suspensions sit in a very different mechanical regime. We therefore do not treat these media as bulk yield-stress fluids, and no analysis in this study relies on a measured $\sigma_{\mathrm{y}}$ or $G^{\prime}$. The consequence for interpretation is important: the confinement a swimming cell experiences here is steric and geometric --- set by grain crowding and by the finite force needed to displace a grain against its drag and its neighbors --- rather than by a macroscopic yield stress that the cell must exceed. Wherever we use the language of caging and cage forces in the main text, it refers to this steric confinement by neighboring grains, by analogy with dense suspensions and soft-particle glasses \cite{book-bonnecaze, seth-2011-natmater, nicolas-2018-rmp}, and not to a measured bulk yield stress in these samples. \\

The single resolved value is nonetheless quantitatively useful. Taken with a grain scale $a_{\mathrm{obs}} \approx$ 10 \SI{}{\text{\micro\meter}}, it gives a cage force $f_{\mathrm{cage}} \sim \sigma_{\mathrm{y}} \, a_{\mathrm{obs}}^{2} \approx$ 3 \SI{}{\text{\pico\newton}}, against a directly measured flagellar force of 26 $\pm$ 5 \SI{}{\text{\pico\newton}} for this organism \cite{boddeker-2020-jrsi}. The mechanical consequence --- that the medium is dense but weak, and therefore remodelable by the swimmer --- is developed in the main text through the remodeling ratio $\mathcal{R} = f_{\mathrm{stall}} / f_{\mathrm{cage}} \approx$ 9. \\

Since viscosity is the only rheological quantity we can determine robustly across the full concentration series, we base the volume fraction assignment on it. We convert $C_{\mathrm{m}}$ to grain volume fraction $\phi$ using the Krieger-Dougherty relation \cite{krieger-1959-transsocrheol} for the relative viscosity of a suspension of deformable spheres, 
\begin{align}
    \frac{\eta_{0}}{\eta_{\mathrm{s}}} = \left( 1 - \frac{\phi}{\phi_{\max}} \right)^{ - [\eta] \, \phi_{\max}},
    \label{eq:si-KD-fit}
\end{align}
where $\eta_{0}$ is the low-shear-rate viscosity of the suspension (here, evaluated for $\dot{\varepsilon}$ = 10 \SI{}{\text{\second\tothe{-1}}}), $\eta_{\mathrm{s}}$ that of the TAP medium, $[\eta]$ = 2.5 the intrinsic viscosity \cite{einstein-1906-annphys, einstein-1911-annphys}, and $\phi_{\max}$ = 0.64 the maximum packing fraction \cite{bernal-1960-nature, torquato-2000-prl}. The measured relative viscosities $\eta_{0} / \eta_{\mathrm{s}}$ is plotted against the resulting $\phi$ in figure \ref{fig:fig-s1}B, together with the Krieger-Dougherty relation (equation \eqref{eq:si-KD-fit}). These $\phi$-estimates are used throughout the manuscript. Since the most concentrated samples are already weakly shear thinning at 10 \SI{}{\text{\second\tothe{-1}}} (figure \ref{fig:fig-s1}A), the viscosity used here is not a true zero-shear value and the procedure therefore yields a conservative (lower) estimate of $\phi$. \\

We are explicit that figure \ref{fig:fig-s1}B documents this mapping rather than independently testing it: because $\phi$ is obtained by inverting equation \eqref{eq:si-KD-fit} from the measured $\eta_{0} / \eta_{\mathrm{s}}$, the data lie on the fitted curve by construction. The panel's purpose is to show the assignment, the range of relative viscosities spanned, and that the highest concentration remains below the $\phi_{\max}$ = 0.64 value used in the Krieger-Dougherty inversion ($\phi$ = 0.568 $<$ $\phi_{\max}$ = 0.64). We note that this is suggestive rather than definitive evidence of being unjammed: $\phi_{\max}$ = 0.64 is the random-close-packing value assumed for rigid spheres, and real microgel grains, being individually compressible, can be packed to somewhat higher effective volume fractions than this before jamming. The near-absence of a measurable yield stress (figure \ref{fig:fig-s1}A) is the more direct evidence that this suspension is crowded but not deeply jammed. \\

Finally, $\phi$ as defined here is a three-dimensional volume fraction for the experiments. The numerical model is two-dimensional and is parametrized by an area fraction $\psi$. The two are not numerically interchangeable and we use distinct symbols throughout to avoid conflation. 

\subsection{Planar versus three-dimensional tracking}

\begin{figure}
    \centering
    \includegraphics[width=\textwidth]{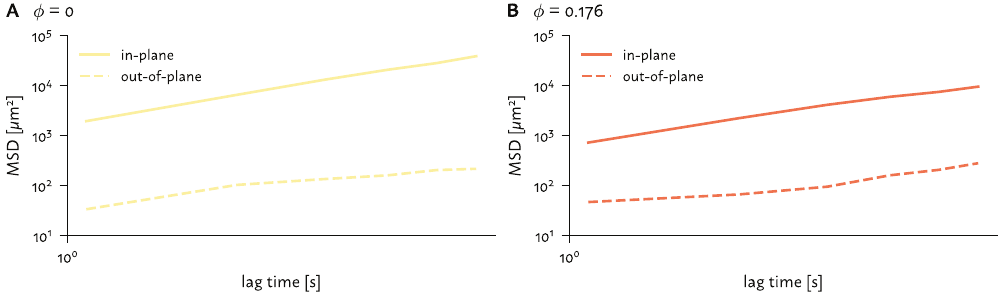}
    \caption{\textbf{Out-of-plane motion contributes negligibly to cell displacement, validating planar tracking}. Mean-squared displacement of individual cells decomposed into in-plane (solid lines) and out-of-plane (dashed lines) components, obtained from three-dimensional tracking of confocal $z$-stacks, for \textbf{A.} the TAP medium ($\phi$ = 0) and \textbf{B.} a microgel packing fraction $\phi$ = 0.176. In both conditions the out-of-plane MSD lies more than an order of magnitude below the in-plane MSD across the full range of lag times. Planar tracking therefore captures the great majority of cell displacement, and the in-plane speeds and hop-and-trap statistics reported in the main text are not biased by projection or by loss of cells from the focal plane. Note that confocal acquisition requires fluorescence excitation and therefore omits the 665 \SI{}{\text{\nano\meter}} long-pass filter used for brightfield imaging. Thus phototactic bias is not suppressed in these measurements; absolute values are not directly comparable to figure \ref{fig:fig-1}C. }
    \label{fig:fig-s2}
\end{figure}

All motility statistics in the main text are extracted from trajectories tracked in a single imaging plane. Because \emph{C. reinhardtii} swims helically and can execute substantial out-of-plane excursions \cite{drescher-2009-rsi, buchner-2021-pnas, mehmood-2025-expfluids, pradipta-2026-prx}, planar tracking could in principle bias the measured speeds, hop lengths, and trapping statistics --- either by systematically foreshortening displacements through projection, or by preferentially losing fast cells that leave the focal plane. That out-of-plane motion matters for this organism is well-established: three-dimensional tracking and holographic flow-field measurements of free-swimming cells resolve helical trajectories and hydrodynamic structure that planar imaging cannot capture \cite{drescher-2009-rsi, buchner-2021-pnas, mehmood-2025-expfluids, pradipta-2026-prx}. We therefore quantified the out-of-plane contribution directly. \\

Cells were tracked in three dimensions from confocal $z$-stacks (13 slices at a step size of 5 \SI{}{\text{\micro\meter}}; Nikon Ti2 inverted microscope equipped with a Nikon C2 point-scanning confocal and a Nikon CFI Plan Apo 10$\times$ NA 0.45 objective, 20 \SI{}{\text{\micro\meter}} pinhole) while they swam through the fluorescently-labeled medium, and the mean-squared displacement (MSD) was decomposed into its in-plane and out-of-plane components. In the TAP medium ($\phi$ = 0, figure \ref{fig:fig-s2}A) and in the microgel at $\phi$ = 0.176 (figure \ref{fig:fig-s2}B), the out-of-plane MSD lies more than an order of magnitude below the in-plane MSD at every lag time.  \\

Note that the confocal $z$-stacks are necessarily acquired with fluorescence excitation, and therefore without the 665 \SI{}{\text{\nano\meter}} long-pass filter used for all brightfield imaging. That filter restricts illumination to wavelengths outside the sensitivity band of the \emph{C. reinhardtii} photoreceptor \cite{hegemann-2008-arplantbiol, kreis-2018-natphys, catalan-2025-natprotoc}, thus suppressing phototactic bias. The confocal measurements cannot do the same, and phototaxis is not negated here. The absolute values in figure \ref{fig:fig-s2} are consequently not directly comparable to the brightfield MSD measurements in figure \ref{fig:fig-1}C, and modest quantitative differences between the two datasets are expected. What transfers is the strong anisotropy between the in-plane and out-of-plane displacements, and it is reproduced at both $\phi$ = 0 and $\phi$ = 0.176 despite the very different microgel structure in the two cases. We therefore expect out-of-plane motion to be negligible in the brightfield measurements as well, and treat the planar tracks as an adequate proxy for the full three-dimensional motility. Moreover, in the confocal microscopy setup, the biasing photo-excitation propagates along the optical axis. So, any phototactic response would act along the out-of-plane direction and, if anything, enhance the out-of-plane displacement; on that reading the measured out-of-plane MSD in figure \ref{fig:fig-s2} is a conservative upper bound on what is obtained under filtered brightfield illumination conditions used in the present work. \\

Two consequences follow. First, in-plane speeds are not materially foreshortened by projection: the planar speed underestimates the full three-dimensional speed by only a few percent, so the speed distributions and the non-monotonic $\phi$-dependence reported in the main text are not projection artifacts. Second, cells remain within the focal plane for the duration of a typical track, so the planar dataset is not biased by selective loss of the most mobile cells --- the concern that would otherwise undermine the hop-and-trap statistics, where a hop terminated by a cell leaving the imaging plane is indistinguishable from a hop terminated by a cage. Planar tracking is therefore an adequate proxy for the three-dimensional motility of these cells under our confinement conditions. 

\subsection{Speed bimodality and non-monotonic swimming speed}

\begin{figure}
    \centering
    \includegraphics[width=\textwidth]{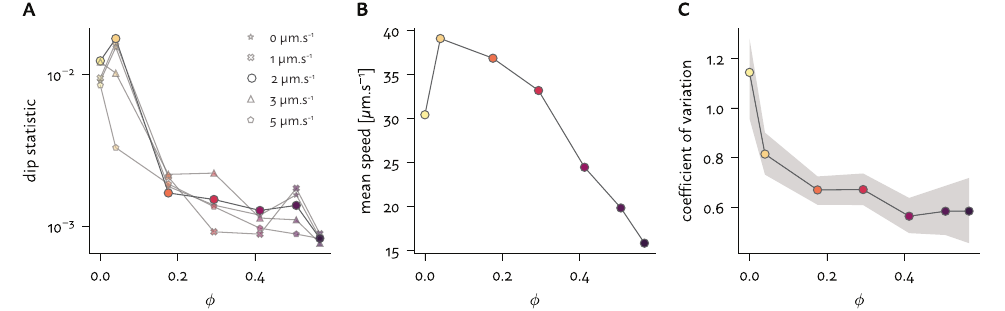}
    \caption{\textbf{Gait bimodality, non-monotonic speed, and persistence narrowing.} \textbf{A.} Hartigan dip statistic versus $\phi$ for different $v_{\min}$ cutoffs. The dip statistic is a model-free effect-size measure of departure from unimodality. The statistic falls by an order of magnitude between the first two gel concentrations ($\phi \neq$ 0) and is approximately flat thereafter. Sweeping across $v_{\min}$ cutoff values show that the multimodal-to-unimodal conclusion is robust to the low-speed cutoff. \textbf{B.} Mean instantaneous speed versus $\phi$, showing the non-monotonic dependence --- a peak in the sparsest gel ($\phi$ = 0.04, not $\phi$ = 0) before a monotonic decline. \textbf{C.} Coefficient of variation of $\tau_{\mathrm{p}}$: the monotonic decline establishes the narrowing of the $\tau_{\mathrm{p}}$ distribution with $\phi$ (shaded band: standard error). Because the coefficient of variation is scale-free, this decline shows the narrowing is relative, not just absolute --- not an artifact of the cells simply slowing down. }
    \label{fig:fig-s3}
\end{figure}

The collapse of the run-tumble speed bimodality (figure \ref{fig:fig-2}) rests on Hartigan's dip test \cite{hartigan-1985-annstat}. We report the dip statistic as a model-free, effect-size measure of multimodality (figure \ref{fig:fig-s3}A). A subtlety here is the low-speed localization floor. At 10 \SI{}{\text{\hertz}}, a displacement at the localization precision ($\approx$ 0.1 \SI{}{\text{\micro\meter}}) corresponds to $\approx$ 1 \SI{}{\text{\micro\meter.\second\tothe{-1}}}, and $\approx$ 13\% of frames in TAP medium fall below this. A naive analysis on the raw speeds latches onto this spike and separates immotile/noise from everything else. Excluding speeds below a cutoff $v_{\min}$ removes this artifact; figure \ref{fig:fig-s3}A shows that the dip-test conclusion --- multimodal at the two lowest $\phi$, unimodal above --- is unchanged over a range of $v_{\min}$, confirming that the collapse of the gait bimodality is not an artifact of this threshold. \\

The mean instantaneous speed is non-monotonic in $\phi$: it peaks for $\phi$ = 0.04 (not $\phi$ = 0), before falling below the free-swimming value at higher $\phi$ (figure \ref{fig:fig-s3}B). This enhancement is traced to the dissolved uncrosslinked polymers in the continuous phase (figure \ref{fig:fig-s5}). \\

The coefficient of variation of the $\tau_{\mathrm{p}}$ distribution (figure \ref{fig:fig-2}E) decreases monotonically with $\phi$ (figure \ref{fig:fig-s3}C), establishing that the persistence-time distribution narrows rather than broadens in relative as well as absolute terms --- the opposite of the broadening expected in a rigid (non-deformable), quenched-disorder landscape \cite{chepizhko-2013-prl, bhattacharjee-2019-natcommun}. Because the coefficient of variation is scale-free, this narrowing cannot be attributed simply to the overall slowing of the cells. 

\subsection{Hop-and-trap controls}

\begin{figure}
    \centering
    \includegraphics[width=\textwidth]{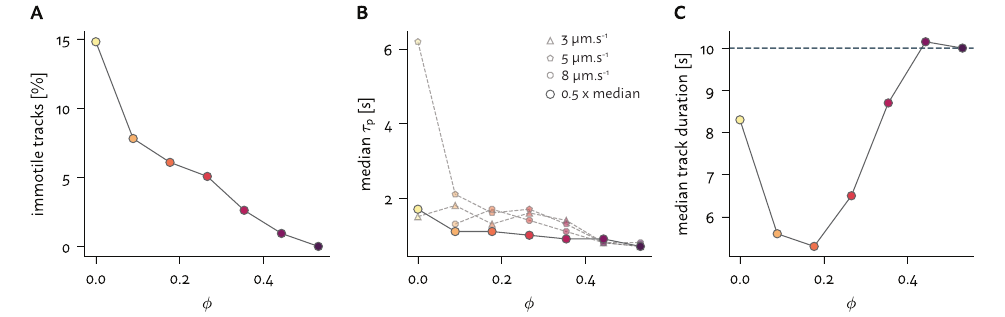}
    \caption{\textbf{Hop-and-trap segmentation controls.} \textbf{A.} Percentage of immotile tracks (median speed $<$ 2 \SI{}{\text{\micro\meter.\second\tothe{-1}}}), falling from 14.8\% at $\phi$ = 0 to 0\% at the highest $\phi$. \textbf{B.} Median trapping time $\tau_{\mathrm{p}}$ versus $\phi$ for fixed speed thresholds (3, 5, 8 \SI{}{\text{\micro\meter.\second\tothe{-1}}}) and for the relative (0.5 $\times$ median) threshold, showing that a fixed threshold sweeping through a shifting speed distribution produces a spurious trend. \textbf{C.} Median track duration versus $\phi$, with the matched 10.0 \SI{}{\text{\second}} observation window marked by the dashed line; the non-monotonic dependence is why unmatched windows would bias conditions differently. }
    \label{fig:fig-s4}
\end{figure}

Three controls underlie the main-text hop-and-trap result (figure \ref{fig:fig-3}), each addressing a way the apparent $\phi$-trend could be an artifact. The fraction of tracks with median speed below 2 \SI{}{\text{\micro\meter.\second\tothe{-1}}} falls from 14.8\% at $\phi$ = 0 to 0\% at the highest $\phi$ (figure \ref{fig:fig-s4}A). This is consistent with --- though not independently confirmed as --- cells settling onto the bottom coverslip over the course of a long recording in TAP medium, while being held in suspension by the microgel dispersion at higher $\phi$; regardless of the precise mechanism, these tracks generate anomalously long apparent traps and are excluded from the hop-and-trap analysis throughout. With a fixed speed threshold, both the median trap time and the hop length show spurious $\phi$-trends (the trap time appears to fall, the hop length appears to rise) --- artifacts of holding the threshold fixed while the underlying speed distribution shifts with $\phi$, not evidence of a genuine trend in either quantity. A threshold referenced to each $\phi$-condition's own median speed removes this (figure \ref{fig:fig-s4}B). Finally, median track duration is non-monotonic in $\phi$ (figure \ref{fig:fig-s4}C), so unmatched windows bias different conditions differently --- at $\phi$ = 0, short tracks cannot contain long traps, biasing its median trap time low. Restricting all conditions to a common 10.0 \SI{}{\text{\second}} window yields the values reported in the main text (figure \ref{fig:fig-3}). The relative threshold introduced here is what makes the segmentation insensitive to the shifting speed distribution, but it also means that the analysis is blind to a uniform rescaling of speed. This is made explicit in figure \ref{fig:fig-s8}A, which constructs a null model that the trapping-time result must be measured against. The broadly distributed hop lengths and residence times parallel those reported for run-and-tumble bacteria near surfaces and in porous media, where residence-time statistics are set by run-to-tumble gait variability \cite{junot-2022-prl}. 

\subsection{Motility in the grain-free supernatant}

\begin{figure}
    \centering
    \includegraphics[width=\textwidth]{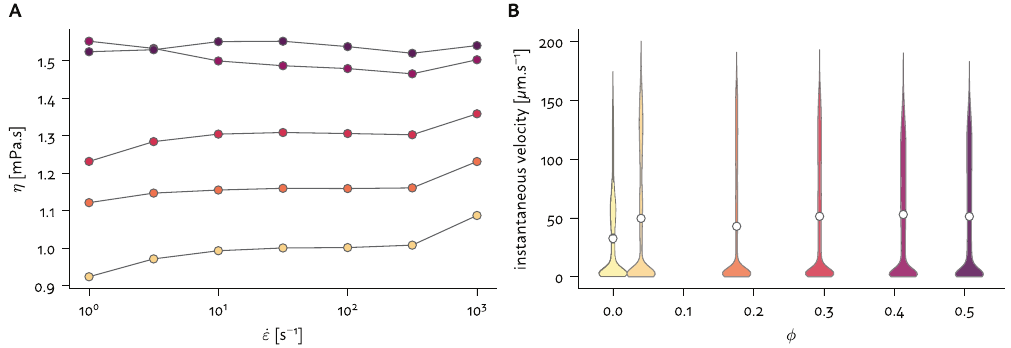}
    \caption{\textbf{The low-$\phi$ speed enhancement is carried by the dissolved, uncrosslinked polymer phase.} \textbf{A.} Shear viscosity $\eta$ versus shear rate $\dot{\varepsilon}$ for the grain-free supernatant corresponding to different volume fractions $\phi$ (marker colors correspond to the corresponding $\phi$ values in panel B). Viscosity rises monotonically with $\phi$, indicating dissolved uncrosslinked polymers in the continuous phase; elastic moduli were below the resolution of our measurement and are therefore not shown. \textbf{B.} Distributions of instantaneous swimming speed in the grain-free supernatant; overlaid discrete markers denote the mean of each distribution, reproducing the increase in mean velocity from $\phi$ = 0 to $\phi$ = 0.04 as seen in the complete gel (figure \ref{fig:fig-2}A). }
    \label{fig:fig-s5}
\end{figure}

To separate the contribution of the discrete microgel grains from that of the continuous phase, the microgel suspension at each $\phi$ was centrifuged at 3000$g$ RCF for 10 \SI{}{\text{\minute}} and the grain-free supernatant retained via decanting. Steady-shear rheometry of the supernatant shows a monotonic rise in viscosity with $\phi$ (figure \ref{fig:fig-s5}A), consistent with a population of dissolved, uncrosslinked Carbopol polymer that is not incorporated into the swollen-grain suspension. Elastic moduli were below the resolution of our measurement and are not reported. \\

Cell tracking in the supernatant reproduces the increase in mean speed from $\phi$ = 0 to $\phi$ = 0.04 of the full gel (figure \ref{fig:fig-2}A) --- establishing that the low-$\phi$ speed enhancement does not require the grain packing and is carried by the dissolved-polymer phase (figure \ref{fig:fig-s5}B). Whether fluid elasticity raises or lowers swimming speed is known to depend sensitively on the swimmer's gait and stroke \cite{teran-2010-prl, thomases-2014-prl}. Our supernatant measurements resolve a rising viscosity but not the associated elastic stresses, so we identify the dissolved, uncrosslinked polymer as the agent of the enhancement without assigning it to a specific viscous or elastic mechanism. 

\subsection{Contact force asymmetry, finite box artifact, and transport map}

\begin{figure}
    \centering
    \includegraphics[width=\textwidth]{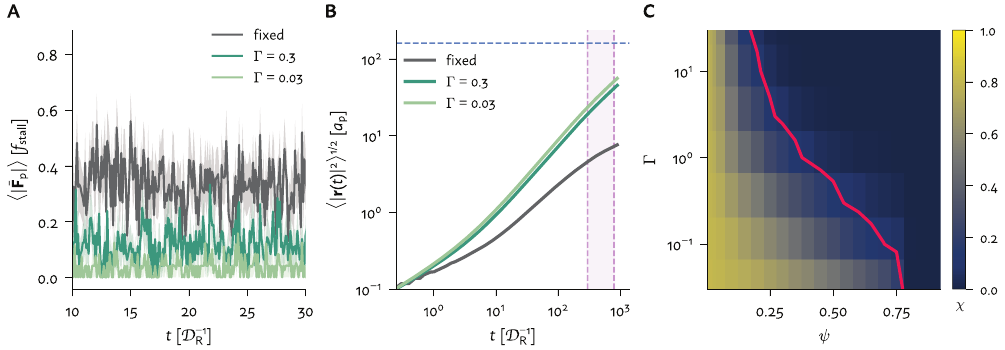}
    \caption{\textbf{Contact force asymmetry, the wall-guarded fit window, and the transport map.} \textbf{A.} Ensemble-averaged magnitude of the contact force the swimmer transmits to obstacles, $\langle \vert \bar{\mathbf{F}}_{\mathrm{p}} \vert \rangle$ (equation \eqref{eq:si-force-cap}), versus time, for fixed and mobile ($\Gamma$ = 0.3 and $\Gamma$ = 0.03) obstacles, at $\psi$ = 0.25 and $\mathrm{Pe}$ = 30 (shaded bands: standard error over $N$ = 96 independent obstacle fields). This is the delivered force, bounded per contact by $f_{\mathrm{stall}}$. Individual records are sequences of brief spikes, so the ensemble average reads as a level set by the fraction of the population in load-bearing contact at any instant. \textbf{B.} Wall-guard illustration: root-mean-squared displacement versus time with the $T_\mathrm{wall}$ cutoff (shaded vertical region) and the $L/2$ bound (horizontal dashed line) marked, at $\psi$ = 0.25, showing the window over which $\mathcal{D}_{\mathrm{eff}}$ is fitted. \textbf{C.} Normalized exploration $\chi = \mathcal{D}_{\mathrm{eff}} / \mathcal{D}_{\mathrm{free}}$ over the $(\psi, \Gamma)$ plane. $\chi$ = 1: free diffusion, $\chi \rightarrow$ 0: complete suppression; the overlaid contour line marks $\chi$ = 0.1. }
    \label{fig:fig-s6}
\end{figure}

The contact force cap $f_{\mathrm{stall}}$ limits the force the swimmer delivers to a mobile obstacle --- this is what decides whether a grain can be pushed --- but it does \emph{not} cap the reaction that a fixed (immobile) obstacle exerts back on the swimmer. There, the swimmer feels the full, uncapped elastic reaction $\sum_{i} \mathcal{K} \, \zeta_{\mathrm{p} i} \, \mathbf{e}_{\mathrm{p} i} \, \Theta (\zeta_{\mathrm{p} i})$ and cannot penetrate an obstacle it is unable to move, whereas the force transmitted to the obstacle $i$ is the capped quantity
\begin{align}
    \bar{\mathbf{F}}_{\mathrm{p} i} = \min (\mathcal{K} \, \zeta_{\mathrm{p} i}, f_{\mathrm{stall}}) \, \mathbf{e}_{\mathrm{p} i} \, \Theta (\zeta_{\mathrm{p} i}) .
    \label{eq:si-force-cap}
\end{align}
This deliberate force asymmetry at the cap represents the cell's finite capacity to do work on the medium while remaining blocked by obstacles it cannot displace; without it, a persistent swimmer would tunnel through immobile grains and the confinement statistics would be spurious. \\

We emphasize first that the two forces are distinct observables and behave oppositely: the delivered force is bounded per contact by $f_{\mathrm{stall}}$ and is what a grain actually receives, while the reaction the swimmer feels is the uncapped Hookean force $\mathcal{K} \, \boldsymbol{\zeta}$, which grows without limit with penetration depth and does not saturate. The text refers to the delivered force throughout when describing figure \ref{fig:fig-s6}A. The consequence of this asymmetry is visible directly in the force the swimmer transmits to the medium (figure \ref{fig:fig-s6}A): the ensemble-averaged magnitude of the delivered force $\langle \vert \bar{\mathbf{F}}_{\mathrm{p}} \vert \rangle$ is bounded by $f_{\mathrm{stall}}$ and remains substantially higher against fixed obstacles, where the swimmer is chronically jammed and delivering its maximum push, than against mobile obstacles ($\Gamma$ = 0.3 and $\Gamma$ = 0.03), where each grain yields and contacts are sparse and short-lived. Because rotational orientation and grain yielding make the individual force records a sequence of brief spikes rather than sustained plateaus, the ensemble average is best read as a steady level set by the fraction of the swimmer population in load-bearing contact at any instant --- higher for a medium that cannot yield, lower for one that can. This is the mechanical counterpart of the exploration result (figure \ref{fig:fig-4}B): the swimmer bears far less load against a mobile medium, which is why it explores it rather than arresting.  \\

A separate, finite-box artifact must also be guarded against when fitting $\mathcal{D}_{\mathrm{eff}}$. The ensemble MSD from the numerical model was fitted for the long-time diffusivity $\mathcal{D}_{\mathrm{eff}}$ over a window $t < T_{\mathrm{wall}}$, chosen so that the root-mean-squared displacement stays below $L / 2$; beyond this bound, reflecting walls fold trajectories back and suppress the apparent $\mathcal{D}_{\mathrm{eff}}$ (figure \ref{fig:fig-s6}B). \\

The full exploration picture behind the reduced presentation in figure \ref{fig:fig-4}C is given in figure \ref{fig:fig-s6}C, which shows the complete $(\psi, \Gamma)$ map of the normalized exploration $\chi$. Increasing obstacle mobility (decreasing $\Gamma$) postpones the $\psi$ at which the localization transition sets in. The boundary at $\chi$ = 0.1, overlaid on figure \ref{fig:fig-s6}C (red contour line), quantifies this directly: the density at which exploration is essentially lost shifts from $\psi \approx$ 0.25 close to the fixed limit ($\Gamma \rightarrow \infty$) to $\psi \approx$ 0.78 at $\Gamma$ = 0.03, a roughly threefold extension in the crowding a swimmer can tolerate, set by obstacle mobility alone. The magnitude of the shift is notable in a wider context: on the lattice, allowing a passive walker to push clusters of obstacles aside moves the localization threshold from the percolation density \cite{bonomo-2026-prl} $\rho_{\mathrm{p}} \approx$ 0.407 to $\rho_{\mathrm{c}} \approx$ 0.91; our continuum, active system shifts its own exploration value over a comparable span. The two thresholds are not directly comparable --- lattice site occupancy versus continuum area fraction, passive versus active, point versus finite-size tracer --- but the correspondence indicates that the ability to displace obstacles relocates the transition. Beyond $\psi \approx$ 0.78, all mobility ratios collapse together to $\chi \rightarrow$ 0, including the most freely displaced case: at these densities the obstacle packing approaches the random close-packing ceiling for hard disks \cite{donev-2004-japplphys, donev-2005-pre} ($\psi \approx$ 0.82--0.84), and a swimmer of bounded propulsive force can no longer displace grains that are mutually jammed against their neighbors. 

\subsection{Strain dependence: flagellar gait sets the sensitivity to soft confinement}

\begin{figure}
    \centering
    \includegraphics[width=\textwidth]{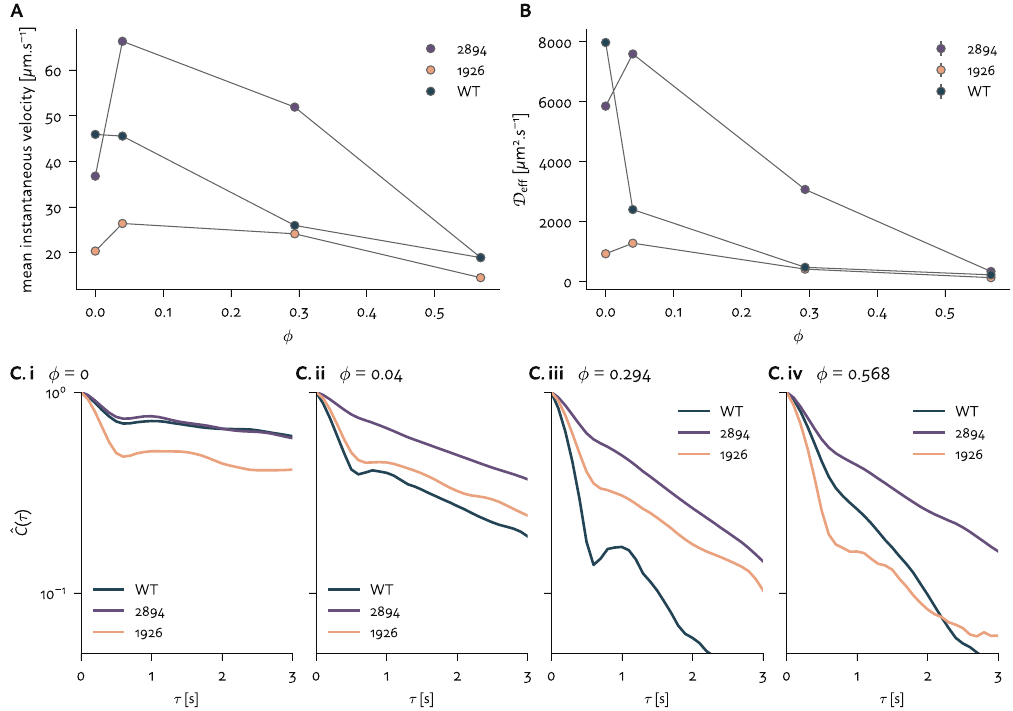}
    \caption{\textbf{Flagellar gait determines the sensitivity of motility to soft confinement.} Wild-type strain 11/32A (WT), asynchronous-flagellar mutant (2894), and uniflagellar mutant (1926) throughout. \textbf{A.} Mean instantaneous velocity and \textbf{B.} effective diffusivity $\mathcal{D}_{\mathrm{eff}}$ versus microgel effective packing fraction $\phi$. Normalized velocity autocorrelation function $\hat{C} (\tau)$ versus lag time $\tau$ for \textbf{C.i.} $\phi$ = 0, \textbf{C.ii.} $\phi$ = 0.04, \textbf{C.iii.} $\phi$ = 0.294, and \textbf{C.iv.} $\phi$ = 0.568. }
    \label{fig:fig-s7}
\end{figure}

To test how flagellar coordination shapes the motility response to soft confinement, we repeated the full tracking and analysis pipeline for two flagellar mutants alongside the wild-type (strain 11/32A): strain 2894, which beats its flagella asynchronously and does not exhibit  phototaxis \cite{zorrilla-2025-prl}, and strain 1926, a uniflagellar mutant propelled by a single flagellum \cite{zorrilla-2025-prl}. All three strains were run through the same microgel packing fraction series and the same analysis (figure \ref{fig:fig-s7}). \\

All strains show the non-monotonic dependence of mean swimming speed with $\phi$ (figure \ref{fig:fig-s7}A). In TAP medium ($\phi$ = 0), the wild-type has the highest effective diffusivity $\mathcal{D}_{\mathrm{eff}}$ (figure \ref{fig:fig-s7}B), consistent with the vigorous breaststroke run-and-tumble gait, while the uniflagellar mutant 1926 has the lowest $\mathcal{D}_{\mathrm{eff}}$ (figure \ref{fig:fig-s7}B). The asynchronous mutant 2894 sits between them in terms of $\mathcal{D}_{\mathrm{eff}}$, echoed in its normalized velocity autocorrelation function $\hat {C} (\tau)$, which decays slowly with a tail beyond 3 \SI{}{\text{\second}} (figure \ref{fig:fig-s7}C.i) --- indicating more persistent runs despite reduced overall exploration. By contrast, 1926 already showed the fastest $\hat{C} (\tau)$ decay in TAP medium free of any microgels (figure \ref{fig:fig-s7}C.i). \\

Adding Carbopol suppresses $\mathcal{D}_{\mathrm{eff}}$ in all three strains, but by markedly different amounts (figure \ref{fig:fig-s7}B). The wild-type $\mathcal{D}_{\mathrm{eff}}$ drops roughly fourfold between $\phi$ = 0 and $\phi$ = 0.04, and continues to fall; the asynchronous mutant 2894 declines far more gradually and retains a significantly higher $\mathcal{D}_{\mathrm{eff}}$ than the wild-type at matched packing fractions, suggesting that asynchronous beating --- lacking coordinated power strokes --- interacts more weakly with the surrounding microgel network. The uniflagellar mutant 1926 shows the steepest relative suppression and the lowest absolute values at high $\phi$, its already-limited propulsion further diminished by confinement (figure \ref{fig:fig-s7}B). \\

The non-monotonic wild-type response --- transiently enhanced reorientation as steric deflection by obstacles increases, before the dense gel suppresses both translational and rotational motion --- is mirrored in the $\hat{C} (\tau)$ distribution at intermediate $\phi$ (figure \ref{fig:fig-s7}C.iii), where only the wild-type develops a pronounced dip near $\tau \approx$ 0.5--1 \SI{}{\text{\second}}, the signature of partial back-scattering (a transient velocity reversal) off obstacles. The mutants do not show the $\hat{C} (\tau)$ dip (figure \ref{fig:fig-s7}C.iii), implying that this non-monotonic reorientation is specific to the interplay between the breaststroke gait and the suspension microstructure rather than a generic swimmer-gel feature. \\

At the highest volume fraction, $\hat{C} (\tau)$ for all strains converges toward faster decay (figure \ref{fig:fig-s7}C.iv), but differences persist: 1926 decays the fastest. This reinforces the conclusion that asynchronous beating is inherently less sensitive to the microgel suspension's constraints, consistent with 2894's persistently slower $\hat{C} (\tau)$ decay across the whole $\phi$-series. \\

Taken together, these results show that flagellar architecture is a key determinant of the confinement response: the wild-type breaststroke is the most sensitive to the gel's mechanical properties, and conclusions drawn from one strain need not generalize across flagellar phenotypes. 

\subsection{Null control for a non-yielding medium}

\begin{figure}
    \centering
    \includegraphics[width=\textwidth]{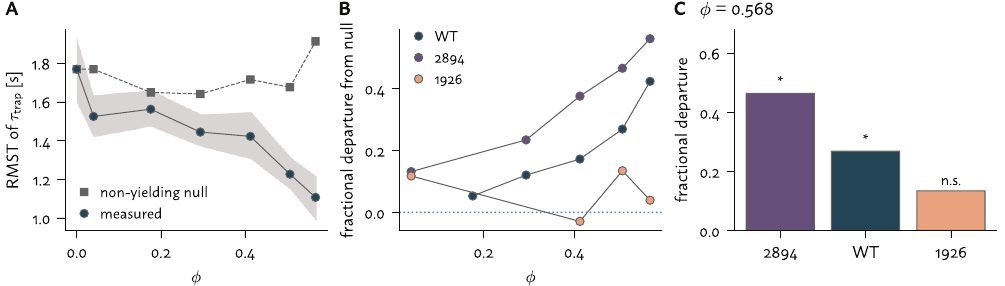}
    \caption{\textbf{Trapping times fall below what a non-yielding medium would produce, and the departure depends on the flagellar apparatus.} \textbf{A.} Restricted mean survival time (RMST) of the trapping-time distribution versus packing fraction $\phi$ for wild-type (11/32A, WT) cells: measured (shaded band denotes bootstrap 95\% confidence interval) and the non-yielding null, constructed by requiring $\phi$ = 0 cells to traverse their own measured paths at the reduced median speed measured at each $\phi$. A larger RMST means cells remain trapped longer. The measured values decrease with crowding while the null increases. \textbf{B.} Fractional departure from the null versus $\phi$ for the WT, 2894, and 1926 mutants. The dimensionless normalization makes strains with different baseline trapping comparable. \textbf{C.} Fractional departure at the highest packing fraction ($\phi$ = 0.568). Asterisks denote departures exceeding the bootstrap 95\% confidence interval for the measurement; n.s.: not significant. The WT and 2894 strains depart significantly from the non-yielding expectation; the uniflagellar 1926 strain does not.  }
    \label{fig:fig-s8}
\end{figure}

The central kinematic claim of figures \ref{fig:fig-3}B, \ref{fig:fig-3}D, and \ref{fig:fig-s4}B  --- that trapping times do not lengthen as the gel densifies, even though the same cells slow markedly --- is a statement about what fails to happen, and therefore requires a quantitative benchmark for what a non-yielding medium would have produced. We construct this benchmark from the data themselves, and then use the flagellar mutants used in figure \ref{fig:fig-s7} to vary the swimmer's capacity to remodel while holding the medium fixed. \\

The segmentation used in figures \ref{fig:fig-3}B, \ref{fig:fig-3}D, and \ref{fig:fig-s4}B is, by construction, insensitive to a pure amplitude rescaling of speed. Because hops and traps are separated by a threshold referenced to each condition's own median speed (figure \ref{fig:fig-s4}B), multiplying every speed in a dataset by any constant rescales the threshold identically and leaves the segmentation frame-for-frame unchanged. We verified this numerically: rescaling all wild-type $\phi$ = 0 speeds by an arbitrary factor reproduces the trapping-time statistics exactly. A medium that merely reduced swimming speed uniformly, without altering the temporal structure of the motion, would therefore be invisible to the analysis --- an invariance measured against such a null would tell us nothing about whether the medium yields. \\

The relevant null is instead time dilation: a rigid medium does not simply lower a cell's speed at fixed dynamics, it makes the cell take longer to do everything, including escaping a cage. We therefore construct a null in which each $\phi$ = 0 cell traverses its same measured path but $1 / \lambda$ times more slowly, where $\lambda (\phi) = \tilde{v} (\phi) / \tilde{v} (0)$ is the measured ratio of median speeds $\tilde{v}$ for that strain. Operationally, each trajectory is re-interpolated onto a time axis stretched by $1 / \lambda$ and re-sampled at the original frame interval, then passed through the identical cleaning, segmentation, and survival-analysis pipeline (see Methods). This null preserves the geometry of the path and the sequence of fast and slow episodes, and changes only the rate at which they are executed --- the behavior expected if the gel impeded the swimmer without yielding to it. Unlike amplitude rescaling it is fully visible to the segmentation, because trap duration is a time-domain quantity whereas the threshold rescales only amplitude. \\

Comparing distributions across strains requires a summary statistic that tolerates heavy censoring --- many trapping episodes are still in progress when a track ends, so their true duration is known only to exceed some lower bound, not measured exactly (``censored'', in survival-analysis terms). Trapping times are censored by track termination in 29--73\% of segments depending on strain and condition, and for the asynchronous mutant 2894, at $\phi$ = 0, the Kaplan-Meier survival function falls only to 0.73 within the matched observation window; so the median used in the main text does not exist. Therefore, we summarize each distribution by its restricted mean survival time (RMST) --- the area under the same Kaplan-Meier curve up to a fixed horizon, here 3.0 \SI{}{\text{\second}} --- which remains well defined even under heavy censoring. We quantify its uncertainty by bootstrapping: resampling the set of tracked cells with replacement many times and taking the spread of the resulting RMST values as the confidence interval \cite{kaplan-1958-jasa, royston-2013-bmcmedresmethodol, uno-2014-jclinoncol}. A larger RMST means cells remain trapped longer. Because RMST and the median are different summaries of the same survival curves, the values here are not directly comparable to the median trapping times in figure \ref{fig:fig-3}D. We report the fractional departure (RMST$_{\text{null}}$ - RMST$_{\text{measured}}$) / RMST$_{\text{null}}$, which is dimensionless and therefore comparable across strains with different baseline trapping --- a positive value means cells escape confinement faster than the non-yielding null predicts (evidence that the medium yields), while a value near zero means the medium behaves as if it simply slowed the cell down without yielding. \\

For WT cells the measured RMST falls monotonically from 1.77 \SI{}{\text{\second}} at $\phi$ = 0 to 1.11 \SI{}{\text{\second}} at $\phi$ = 0.568, while the null --- cells executing their own free-swimming paths at the measured reduced speed --- remains between 1.64 \SI{}{\text{\second}} and 1.91 \SI{}{\text{\second}} (figure \ref{fig:fig-s8}A). Across the conditions for which the null is well-defined, the fractional departure grows monotonically (figure \ref{fig:fig-s8}B). The measured RMST falls with crowding while the null's predicted RMST does not (figure \ref{fig:fig-s8}A) --- cells in the densest gel escape local confinement faster, in absolute time, than they do in free suspension, whereas a medium that slowed them down would have trapped them for longer. The absence of trap broadening reported in figure \ref{fig:fig-3}D is therefore not a null result but a measured departure from the rigid-porous-medium expectation. \\

The same analysis applied to the flagellar mutants tests the mechanism by varying the swimmer rather than the medium. The remodeling ratio $\mathcal{R} = f_{\mathrm{stall}} / f_{\mathrm{cage}}$ can be changed by either altering the medium deformability --- not possible in our Carbopol suspension, where grain mobility cannot be switched off --- or by altering the swimmer, and the mutants used for figure \ref{fig:fig-s7} provide the latter. Restricting attention to well-defined nulls, the fractional departure rises monotonically with crowding for both biflagellate strains WT and 2894 (figure \ref{fig:fig-s8}B). For the uniflagellar mutant 1926 it does not: the departure remains close to zero across the entire series, and is not significant at any of the three highest packing fractions (figure \ref{fig:fig-s8}B). Within measurement error, the uniflagellar 1926 mutant behaves in the densest gels as though the medium did not yield to it. \\

This ordering is not explained by swimming speed. At the highest packing fraction ($\phi$ = 0.568), the mean instantaneous speeds of the three strains are almost identical (figure \ref{fig:fig-s7}A). The fractional departures at the same condition, however, show a much higher degree of contrast (figure \ref{fig:fig-s8}C). What tracks the effect is therefore the flagellar apparatus --- two flagella versus one --- rather than how fast the cell swims.  \\

What this control establishes, and what it does not, is worth stating plainly. It establishes that the motion's temporal structure is not simply slowed down: something actively shortens the below-threshold episodes as crowding increases, and this depends on the swimmer's flagellar apparatus. It does not by itself prove grain displacement is the cause --- a cell that reoriented more frequently could in principle also find an exit sooner. Identifying remodeling as the mechanism rests on direct confocal observation (figures \ref{fig:fig-3}A.i and \ref{fig:fig-3}A.ii, and movie SM2) and on the force estimate $\mathcal{R} \approx$ 9; this control adds kinematic support for that mechanism.

\subsection{Grain compliance alone also partially relieves suppression}

\begin{figure}
    \centering
    \includegraphics[width=\textwidth]{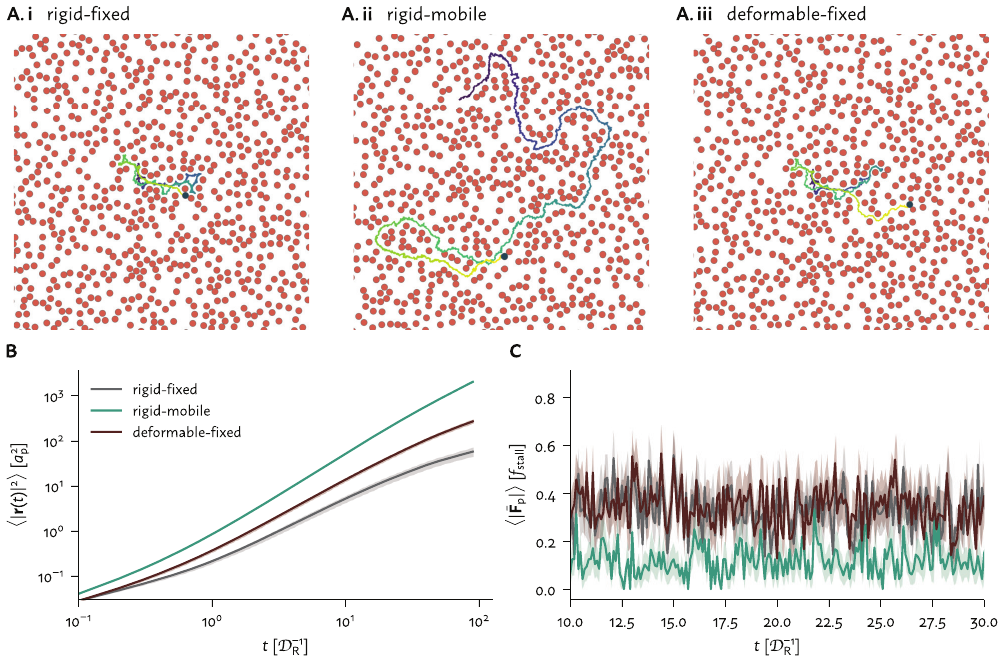}
    \caption{\textbf{Grain compliance alone partially relieves the suppression a rigid medium imposes.} Simulation snapshots at $\psi$ = 0.25, with the same disordered obstacle field (red disks), for \textbf{A.i.} rigid-fixed, \textbf{A.ii.} rigid-mobile ($\Gamma$ = 0.3), and \textbf{A.iii.} deformable-fixed obstacles. The swimmer (green disk) trajectory is overlaid on the obstacle field and the trajectory color encodes elapsed time, increasing monotonically along each path. Each snapshot is zoomed to the vicinity of the swimmer's starting position and does not show the entire obstacle field. See movie SM4 for the corresponding simulation movies. \textbf{B.} Ensemble mean-squared displacement and \textbf{C.} ensemble-averaged magnitude of the contact force the swimmer transmits to obstacles, for the three cases shown in panels A.i, A.ii, and A.iii (shaded bands: standard error over 24 independent obstacle fields); legend shared with panel B. The force record in panel C is an illustrative time window from a much longer dataset; individual records are sequences of brief spikes, so the ensemble-averaged reads as a level set by the fraction of the population in load-bearing contact at any instant. Note that the deformable-fixed case delivers the most force while exploring less than rigid-mobile: delivered force measures work done on the medium, not confinement.  }
    \label{fig:fig-s9}
\end{figure}

The main-text model isolates obstacle \emph{mobility} by treating grains as rigid disks that translate. To confirm that the complementary mechanism --- grain size \emph{deformation} with no translation --- also relieves suppression, we ran a control in which obstacles are held fixed in place but allowed to change in size under contact, so the swimmer can indent a grain it cannot move. \\

A swollen Carbopol microgel grain is not an incompressible disk: it is a solvent-filled polymer network that deforms under load, with the deformation rate-limited by solvent draining through the network \cite{book-bonnecaze, book-fernandeznieves, mattsson-2009-nature, gnan-2019-natphys, nikolov-2020-pnas}. Its size therefore responds to load viscoelastically --- over a finite relaxation time $\tau_{\mathrm{sq}}$ set by the balance between network elasticity and poroelastic drainage. We model this by giving obstacle $i$ a load-dependent radius $a_{i}$ that relaxes toward a compressed target,
\begin{subequations}
    \begin{align}
        \tau_{\mathrm{sq}} \frac{\mathrm{d} a_{i}}{\mathrm{d} t} &= a_{i}^{\mathrm{target}} (\bar{F}_{\mathrm{p} i}) - a_{i} , \\
        a_{i}^{\mathrm{target}} &= a_{\mathrm{obs}} \left( 1 - \min \left( \frac{\bar{F}_{\mathrm{p} i}}{F_{\mathrm{y}}} , 1 - \frac{a_{\min}}{a_{\mathrm{obs}}} \right) \right) ,
    \end{align}
\end{subequations}
where $\bar{F}_{\mathrm{p} i} = \min (\mathcal{K} \, \zeta_{\mathrm{p} i}, f_{\mathrm{stall}})$ is the (capped) magnitude of the swimmer's contact force on grain $i$ (as in equation \eqref{eq:si-force-cap}), $F_{\mathrm{y}}$ sets the compliance, and $a_{\min}$ is the floor radius. The relaxation is reversible: on release, $a_{i}$ re-swells to $a_{\mathrm{obs}}$ over the same time. As in the main-text model, the swimmer feels the full, uncapped elastic reaction force from the immobile obstacle, evaluated at the grain's current radius, and cannot tunnel through it; only the radius of the grain responds, and only to the capped contact force. Grain centres $\{ \mathbf{c}_{i} \}$ are held fixed; all other parameters match the ones used for the simulations reported in the main text. \\

The amount of relief deformation provides depends genuinely on $\tau_{\mathrm{sq}}$: a grain with a long ``memory'' stays compressed briefly after the swimmer begins to pass, holding the transient gap open long enough for the swimmer to slip through, whereas a grain that re-swells instantly closes the gap behind the retreating contact. Deformation therefore relieves confinement through a competition between how long the swimmer sustains a contact and how quickly the grain recovers --- not through an instantaneous geometric shrink (the confocal micrographs in figures \ref{fig:fig-3}A.i and \ref{fig:fig-3}A.ii, and movie SM2, point toward finite swimmer-grain contact and post-contact recovery timescales). We do not derive $\tau_{\mathrm{sq}}$ from the microgel's poroelastic parameters; we fix it at $\tau_{\mathrm{sq}} = 0.2 / \mathcal{D}_{\mathrm{R}}$, of order the swimmer-grain contact duration. \\

At matched area fraction ($\psi$ = 0.25) and matched persistence ($\mathrm{Pe}$ = 30), the trajectories (figures \ref{fig:fig-s9}A.i, \ref{fig:fig-s9}A.ii, and \ref{fig:fig-s9}A.iii, and movie SM4) and the ensemble mean-squared displacements (figure \ref{fig:fig-s9}B; composed of 24 independent realizations per condition) show the same ordering: the deformable-fixed (obstacles are fixed in position but can undergo compression due to contact forces) medium recovers a fraction of the exploration lost in the rigid-fixed limit (obstacles are fixed in position and their radius cannot be altered by contact forces), but not as significantly as the rigid-mobile case (obstacles can be displaced by contact forces but their radius cannot be altered; here grain mobility $\Gamma$ = 0.3). Grain deformability and grain mobility are therefore each independently sufficient to relieve confinement, but to different degrees for the particular case shown in figure \ref{fig:fig-s9}. \\

The contact-force records (figure \ref{fig:fig-s9}C) show that the two mechanisms achieve this very differently, and that the force delivered to the medium is not a proxy for confinement. The deformable-fixed case delivers the most force, yet it explores less than the rigid-mobile case. A deformable grain that cannot translate must be indented for the swimmer to pass; so the swimmer sustains deep, near-stall presses. A mobile grain instead yields by moving aside, and the contact breaks before a large load develops. Exploration is thus purchased far more cheaply by mobility than by deformation: rigid-mobile achieves the highest exploration while transmitting the least force, whereas deformable-fixed does the most mechanical work on the medium for an intermediate gain in transport. The rigid-fixed case, which delivers an intermediate force and gains nothing, completes the picture --- a swimmer pressing on walls that neither move nor deform. This asymmetry motivates a future systematic study spanning grain compliance $F_{\mathrm{y}}$, response time $\tau_{\mathrm{sq}}$, and mobility $\Gamma$, in which the transport gained per unit force delivered could be mapped as a function of all three. 

\subsection{Supplementary movies}

\textbf{Movie SM1:} Time-lapsed brightfield micrographs of wild-type (11/32A) \emph{C. reinhardtii} swimming in TAP medium ($\phi$ = 0), showing extended weakly-curved runs punctuated by reorientations, and in microgel suspension at $\phi$ = 0.568, showing tortuous, spatially confined trajectories. \\[-0.5ex]

\textbf{Movie SM2:} Time-lapsed dual-channel confocal micrographs of swimming cells (green) within the labeled microgel medium (red), at $\phi$ = 0.176 and $\phi$ = 0.568, showing local displacement of the microgel around the advancing cell and its elastic recovery. \\[-0.5ex]

\textbf{Movie SM3:} Numerical simulations at $\psi$ = 0.25, with fixed obstacles --- the active particle is confined to small repeatedly re-traced region, and with mobile obstacles ($\Gamma$ = 0.3 and $\Gamma$ = 0.03) on the same obstacle field --- the active particle displaces grains and explores a much larger area. \\[-0.5ex]

\textbf{Movie SM4:} Numerical simulations at $\psi$ = 0.25 for rigid-fixed, rigid-mobile ($\Gamma$ = 0.3), and deformable-fixed obstacles on a common obstacle field. 


\newpage

\FloatBarrier

\bibliographystyle{unsrt}
\bibliography{chlamy-confinement}


\end{document}